\documentclass[aps,pra,twocolumn,tightenlines,longbibliography,superscriptaddress,doublespace,nofootinbib]{revtex4-2}
\usepackage{subcaption}

\usepackage{graphicx}
\usepackage{setspace,color,appendix,physics,wasysym,xcolor,xparse}
\usepackage{appendix} 
\usepackage[english]{babel}
\usepackage{amsmath,amssymb,amsfonts,mathtools}
\usepackage{mathtools}

\definecolor{brickred}{rgb}{0.8, 0.25, 0.33}
\definecolor{celestialblue}{rgb}{0.29, 0.59, 0.82}
\definecolor{cornflowerblue}{rgb}{0.39, 0.58, 0.93}
\definecolor{denim}{rgb}{0.08, 0.38, 0.74}
\definecolor{armygreen}{rgb}{0.29, 0.33, 0.13}
\definecolor{cardinal}{rgb}{0.77, 0.12, 0.23}
\definecolor{carnelian}{rgb}{0.7, 0.11, 0.11}
\definecolor{armygreen}{rgb}{0.29, 0.33, 0.13}

\usepackage[colorlinks=true, urlcolor=black, citecolor=blue, linkcolor=red, citebordercolor={1 0 0}, linkbordercolor={0 0 1}]{hyperref}

\usepackage{verbatim,multirow}
\usepackage[T1]{fontenc}
\usepackage{lmodern}
\usepackage{pxfonts}
\usepackage{orcidlink}

\begin{document}

\title{Continuous variable payload transmitter for the SPOQC mission}
\author{Vinod N. Rao\,\orcidlink{0000-0003-0364-1834}}
\email{vinod.rao@york.ac.uk}
\affiliation{School of Physics, Engineering \& Technology and York Centre for Quantum Technologies, University of York, YO10 5FT York, U.K.}
\author{Killian Murphy\,\orcidlink{0000-0001-5665-8248}}
\affiliation{IT Services, University of York, YO10 5FT York, U.K.}
\author{Timothy Ayers}
\affiliation{School of Physics, Engineering \& Technology and York Centre for Quantum Technologies, University of York, YO10 5FT York, U.K.}
\author{Fadi Ahwal\,\orcidlink{0000-0002-5512-0275}}
\affiliation{School of Physics, Engineering \& Technology and York Centre for Quantum Technologies, University of York, YO10 5FT York, U.K.}
\author{Timothy P. Spiller\,\orcidlink{0000-0003-1083-2604}}
\affiliation{School of Physics, Engineering \& Technology and York Centre for Quantum Technologies, University of York, YO10 5FT York, U.K.}
\author{Rupesh Kumar\,\orcidlink{0000-0002-6813-6401}}
\email{rupesh.kumar@york.ac.uk}
\affiliation{School of Physics, Engineering \& Technology and York Centre for Quantum Technologies, University of York, YO10 5FT York, U.K.}

\begin{abstract}
The continuous variable (CV) quantum key distribution (QKD) scheme enables two parties to share a secret key separated by a certain distance. Here we propose, design, build, test and show results of a CV payload transmitter, intended to demonstrate downlink based QKD from a CubeSat in low earth orbit. This is part of the SPOQC mission by Quantum Communications Hub, UK which was successfully launched on March 2026. 
\end{abstract}

\maketitle

\section{Introduction \label{sec:intro}}

The necessity for secure communication across continental distances are more today than ever before. The aspect of using quantum systems for communication has been explored since last four decades \cite{bennett1984quantum, ekert1991quantum, bennett1992quantum2}. The space based quantum communication is the recent area where various missions have been executed as well as many are underway. Flagship missions such as Micius \cite{lu2022micius}, Jinan-1 \cite{li2025microsatellite} have successfully demonstrated space-to-ground entanglement distribution and quantum key distribution (QKD) at intercontinental distances. At the CubeSat scale, experiments like SOCRATES \cite{takenaka2017satellite}, SpooQy-1 \cite{villar2020entanglement} and planned missions like QUBE \cite{knips2022qube}, SpeQtre \cite{Speqtre}, QEYSSat \cite{scott2020qeyssat}, etc \cite{armengol2008quantum, jennewein2014nanoqey, balakier2025high, hiemstra2025european} are extending these capabilities into cost effective platforms. Here we provide the details of the quantum payload designed for the UK's Satellite Platform for Optical Quantum Communications (SPOQC) mission \cite{SPOQC}, to be launched in early 2026. The continuous variable (CV) based transmitter/payload for downlink-based QKD is designed, built and tested whose results are provided here.

QKD enables two distant parties in sharing a secret key, whose security is proven by quantum mechanics \cite{scarani2009security, lo2014secure, jain2022practical}. This is distinct from classical cryptographic protocols, where the security comes from the assumption of irreversibility of a mathematical function \cite{zurek2003decoherence, henderson2001classical, portmann2022security}. The primary trade off between these two would be optimal usage of resource vs necessity of security. The classical cryptographic protocols are resource efficient, but offer security based on computational complexity. In contrast, QKD offers information theoretic security at the cost of utilising non-classical correlations. QKD comes in two flavours: discrete variable and continuous variable (CV) \cite{zhang2018large, pirandola2020advances}. The former one involves utilising quantum systems of finite dimensional Hilbert space for encoding, where as the latter (ideally) involves a system with infinite dimensionality of Hilbert space.

Continuous variable quantum key distribution (CV-QKD) has emerged as a promising technology for information-theoretic secure communications, leveraging the quantum properties of coherent states and standard telecom components to enable high rate key generation over optical channels. While fibre based CV-QKD has matured significantly, extending secure quantum links to global scales remains challenging due to the exponential loss in terrestrial fibres. Free space quantum links from satellites in Low Earth Orbit (LEO) offer a compelling pathway toward worldwide quantum networks, enabling long distance key distribution between ground stations separated by thousands of kilometres. Various theoretical frameworks have been proposed for both uplink and downlink based quantum communication. Additionally, various factors affecting the key rates have been investigated as well. Several space based quantum communication is experimentally proven, with more missions planned for future. Complementarily, numerous optical ground stations are developed, intended solely for quantum communication. Demonstrations such as those enabled by the Micius mission have validated the feasibility of satellite based quantum communications, primarily using discrete-variable protocols. However, the realization of space borne CV-QKD remains comparatively unexplored, particularly within the size, weight, and power (SWaP) constraints of CubeSat platforms.

CubeSats provide a cost-effective and rapidly deployable platform for space quantum experiments, but they impose stringent requirements on payload miniaturization, thermal stability, optical alignment, and pointing performance. CV protocols are especially sensitive to excess noise, phase stability, and detection efficiency, making the translation from laboratory systems to space qualified hardware non-trivial. In the LEO environment, additional impairments arise from atmospheric turbulence, Doppler effects, platform vibration, and pointing jitter, all of which can degrade the achievable secret key rate. Consequently, a systematic approach to the designing, building, testing, and verification of a dedicated CV-QKD payload is essential prior to in-orbit demonstration.

This work presents the end to end development of a CubeSat compatible payload designed to demonstrate CV-QKD from LEO. The study addresses the complete engineering chain, including optical architecture, coherent state preparation, homodyne detection subsystem, timing and synchronization, and fine pointing considerations. Additionally, the details about computing architecture \& electronic systems involved in the payload are also provided. Laboratory validation and hardware in-the-loop testing are used to verify system performance against theoretical expectations and mission requirements.

The contributions of this paper are twofold. First, we propose a compact CV-QKD payload architecture compatible with CubeSat SWaP constraints while maintaining the phase stability required for coherent detection. Second, we provide a verification framework that links component level measurements to expected in-orbit behaviour, thereby reducing mission risk and informing future space quantum network designs. Additionally, a note on development and validation of performance testing of electronics in the CVP is also made.

CV based quantum communication usually involves preparing and measuring of Gaussian states, with either Gaussian modulation or discrete modulation \cite{sayat2024satellite, usenko2026continuous}. Though non-Gaussian states can be used for certain cryptographic protocols \cite{ra2020non, lvovsky2020production, jeng2025entanglement}, the efficient quantification of extractable information and validation by experimental implementation remains to be a challenge \cite{straka2018quantum, jain2022practical}. CV quantum cryptography generally entails using coherent, squeezed or thermal states as quantum systems for encoding and coherent detection (homodyne or heterodyne) for decoding. CV-QKD using coherent states has been experimentally implemented both in fibre as well as free space, in addition to it being proven to be secure.

By bridging the gap between theoretical CV-QKD protocols and space-qualified implementation, this work advances the feasibility of CV quantum communications from small satellite platforms and supports the development of scalable global quantum key distribution infrastructures. The CV payload (CVP) for SPOQC is $\approx 3/4$U unit, that is designed to transmit Gaussian modulated coherent states (GMCS). In a GMCS protocol, a train of weak coherent pulses are amplitude and phase modulated which forms the encoding. The decoding is performed by homodyne or heterodyne measurement on the received coherent states. This forms the quantum part of the protocol, which is followed by classical post processing of error estimation, error correction and privacy amplification. Since we are transmitting the local oscillator (LO) with the signal, the consequences of Doppler effect and relative phase drift inside Bob's station are negligible. The LO is transmitted along with the signal via polarisation multiplexing.

\begin{figure}[ht]
\centering
\includegraphics[width=\linewidth]{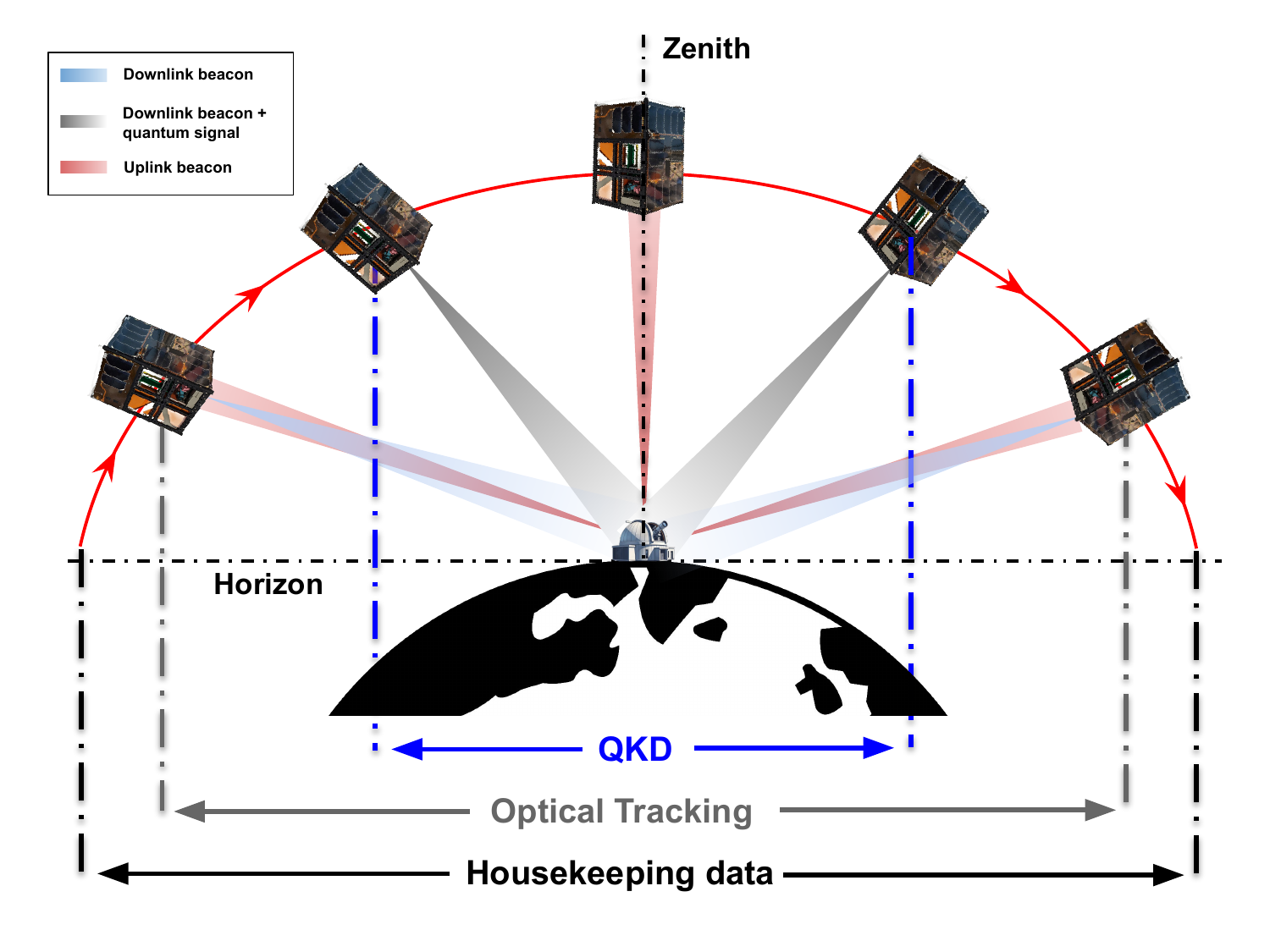}
\caption{The operational concept of the CVP in SPOQC is shown. In the optical pass, the CVP will be powered on prior to the SPOQC reaching the horizon of the OGS. At $30^\circ$ elevation angle, SPOQC and OGS initiate closed loop optical beacon laser tracking, CVQKD signal and LO transmission starts at $60^\circ$. At $120^\circ$ elevation angle, CVP terminated the QKD link and initiate the preparation for data reconciliation for the RF pass for classical communication and shutdown the payload. Throughout the all operational stages, house keeping and monitoring data will be recorded.}
\label{fig:sat_track}
\end{figure}

A brief operational concept of the payload is shown in Fig. \ref{fig:sat_track}, with the various operations of the payload being designed for a range of different elevation angles (and thus different time duration). Quantum and classical communications are separated as optical pass and Radio Frequency (RF) pass. A configuration file defines the operational sequence of the CVP is pre-shared before the optical pass. An optical pass and multiple RF passes are required for the generation of the key.

This paper is arranged as follows. After a brief introduction in Sec. \ref{sec:intro}, we outline the design details in Sec. \ref{sec:design}. This is divided into various subsections, where each unit is detailed. The verification and validation of the performance of CVP is discussed in Sec. \ref{sec:test}. Additionally a brief note on environmental test procedures, including the integration with the CubeSat, are provided. Finally discussions \& conclusions are in Sec. \ref{sec:disc}.

\section{Design and operation of CVP \label{sec:design}}

\subsection{System Overview \label{subsec:system}}

\begin{figure}
\centering
\begin{subfigure}{0.5\textwidth}
\includegraphics[width=0.95\linewidth]{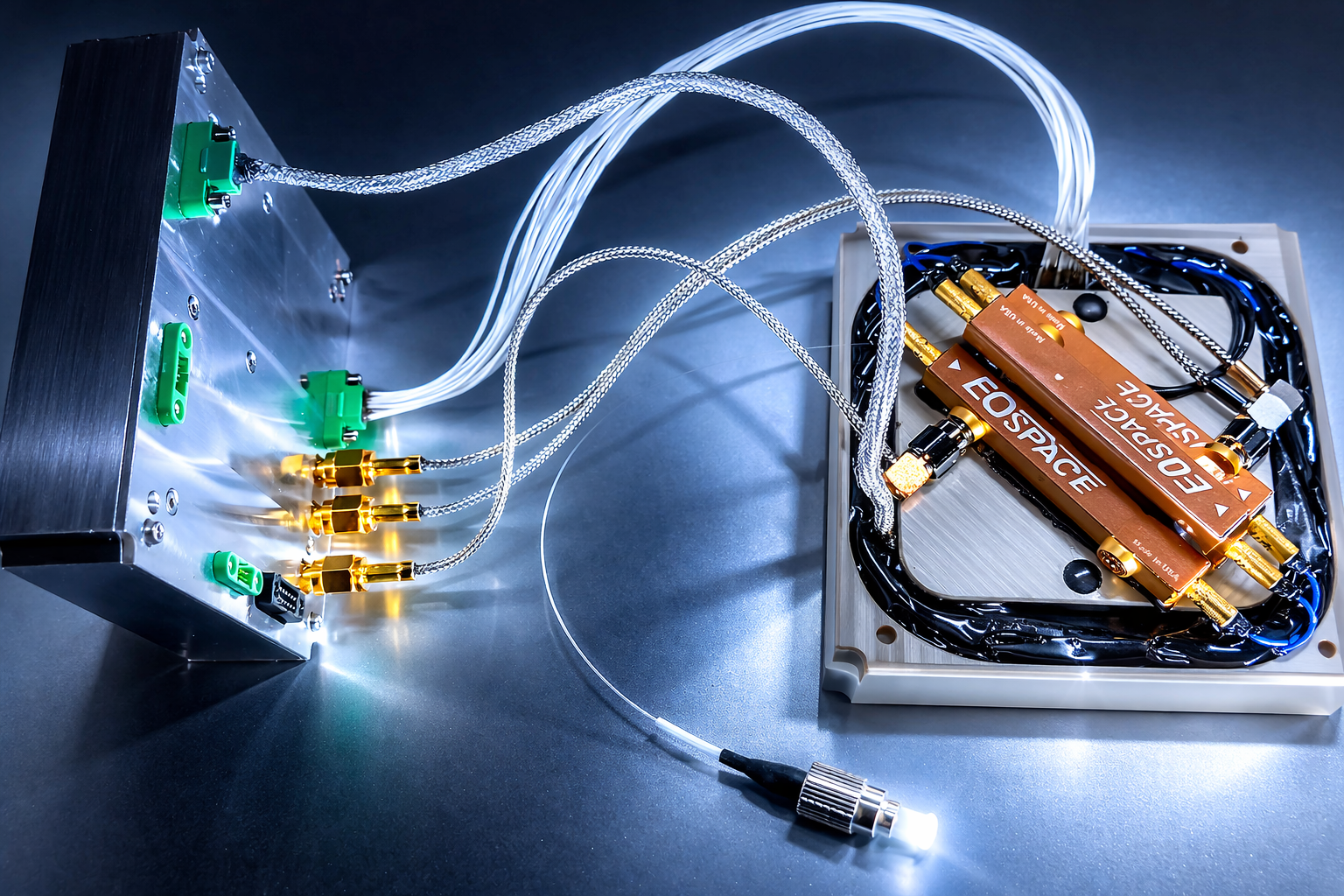}
\caption{}
\label{fig:CVP_pic} 
\end{subfigure}
\medskip
\begin{subfigure}{0.5\textwidth}
\includegraphics[width=0.95\linewidth]{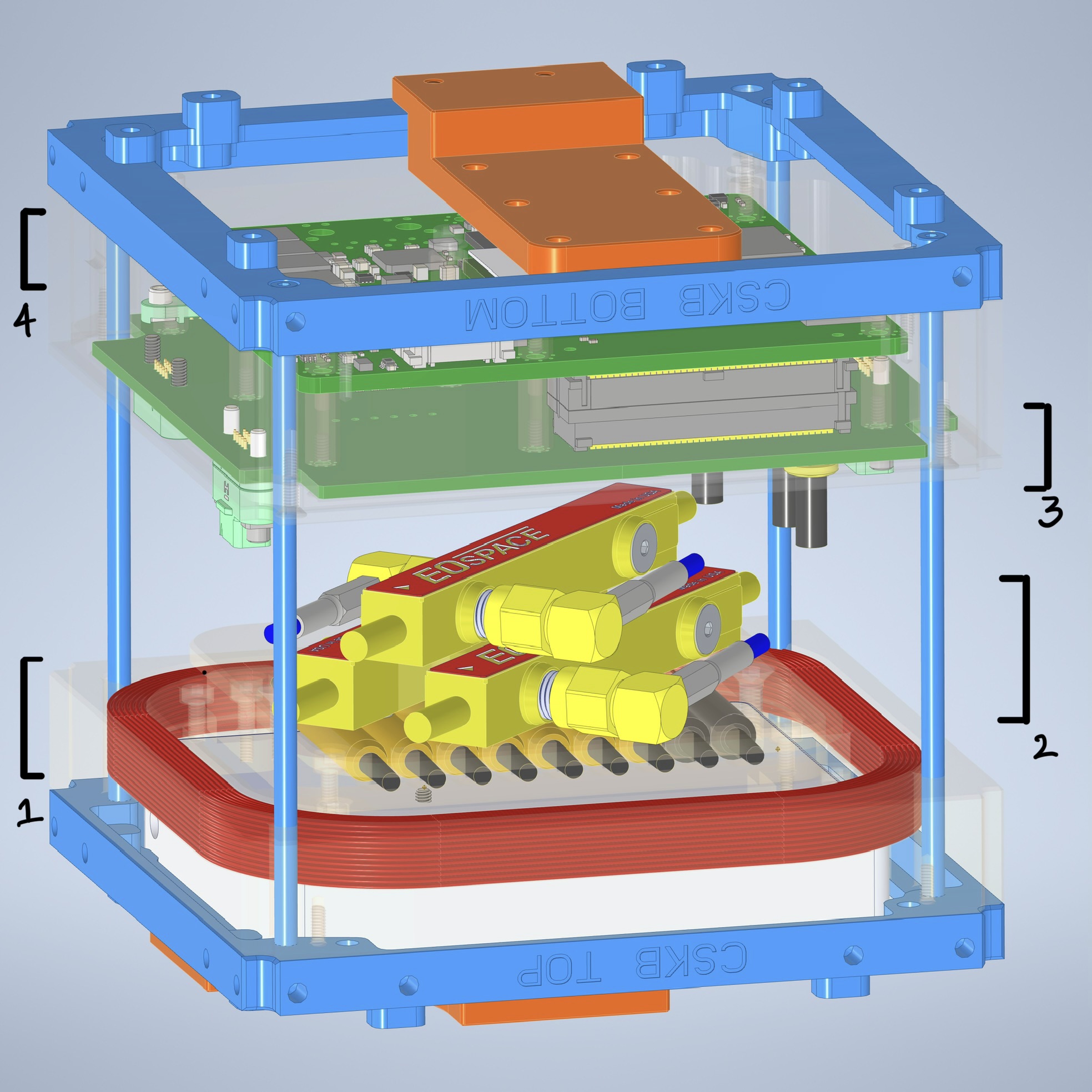}
\caption{}
\label{fig:CVP_3D}
\end{subfigure} 
\caption{The photo of the CVP (Fig. \ref{fig:CVP_pic}) and the 3D image of the same (Fig. \ref{fig:CVP_3D}) are given here. In Fig. \ref{fig:CVP_3D}, the four units are - $1 \rightarrow$ laser; $2 \rightarrow$ optical components; $3 \rightarrow$ SBC; $4 \rightarrow$ daughter board.}
\label{fig:CVP}
\end{figure}

The CVP is designed to provide a proof of concept prepare and measure CV-QKD transmitter, based on GMCS, along with a ground based homodyne receiver. The design is constrained to a \textbf{1U CubeSat form factor ($\mathbf{10} \times \mathbf{10} \times \mathbf{10}~\text{cm}^3$)} and optimized for low SWaP while maintaining the phase and amplitude stability required for efficient performance as shown in Fig. \ref{fig:CVP}. The SWaP details are as follows: size $\leq 3/4~\text{U}$; weight $< 1~\text{kg}$; power (optical pass) $\leq 15~\text{W}$; power (RF pass) $\leq 5~\text{W}$. The image of the CVP designed for SPOQC mission is shown in Fig. \ref{fig:CVP_pic}, and the four units of the CVP are marked in Fig. \ref{fig:CVP_3D}. The working of CVP is supported by a power source and a central mission computer, which are not considered here.

The payload architecture follows the standard CV-QKD signal chain. We have a narrow linewidth laser source, following which optical power is split into signal and LO paths. Gaussian modulation of optical quadratures are performed for signal pulses. Subsequently, after the signal is multiplexed with LO, the beam is sent through the transmitter telescope (after beam expansion). Within the CubeSat, the CVP is compartmented into two main modules: optics and electronic, each subdivided into two units further. Optical module has laser and optical elements. The optical elements include modulator units for signal modulation and a homodyne detection setting for quantum random number generation (QRNG). The electronic module has single board computer (SBC) for data processing and communications and a daughter board corresponding to electric signal generation for signal modulation and QRNG data conversion. In addition to most of the elements being space qualified, the components need to be compact and must be within the allocated weight limit as well as optimised for electrical power usage.

The presented CVP demonstrates that GMCS based CV-QKD can be implemented within stringent SWaP constraints through careful photonic integration, transmitted LO architecture, and closed loop modulator stabilization. The design provides a practical pathway toward space borne CV quantum communications and scalable global quantum networks. Fitting the CV-QKD transmitter within 1U requires meticulous planning and designing in addition to succinct integration. Additionally, the key parameters of the CVP after its integration with the SPOQC CubeSat is given in Tab. \ref{tab:para}.

\begin{table}[htbp]
\centering
\footnotesize
\setlength{\tabcolsep}{4pt}
\begin{tabular}{|l|c|}
\hline
\textbf{Parameter} & \textbf{Details} \\
\hline
\hline
CVP source rate & $2~\text{MHz}$ \\
CVP signal wavelength & $1550.12~\text{nm}$ ($\pm 5~\text{nm}$) \\
CVP signal pulse width & $2~\text{ns}$ \\
CVP signal configuration & Downlink \\
CVP signal type & Pulsed laser source \\
CVP signal power & $<1~\text{pW}$ \\
Avg LO power & $< 1~\mathrm{W}$ \\
Multiplexing scheme & Polarisation (\& time) multiplexing \\
Beam divergence & $22.5~\mu\text{rad}$ \\
Degree of polarisation & $99.51\%$ \\
Azimuth & $96.46\%$ \\
Ellipticity & $98.17\%$ \\
Polarisation compensation & Not required (ideal) \\
\hline
\end{tabular}
\caption{Key parameters of the CV payload in SPOQC mission.}
\label{tab:para}
\end{table}

\subsection{Optical module \label{subsec:opt_mod}}

The optical module consists of compact laser module and an optical fibre components module. The Erbium doped fibre based laser module (BKTel) generates $1550.12~\text{nm}$ wavelength pulses of $2~\text{ns}$ width. The monitoring fibre port from the laser module provides $2~\text{mW}$ peak power while the strong fibre port with digital intensity control gives out $250~\text{W}$ maximum peak power at $2~\text{MHz}$ repetition rate, which serves as the TLO. The laser module comes with digitally monitoring features such as temperature, status and alarm. The design of the optical module in the CVP is shown in Fig. \ref{fig:design}.

The signal from the first beam splitter is divided into two: one for generating GMCS while the other for QRNG purposes. We use two amplitude modulators of total $60~\text{dB}$ dynamic range, and a phase modulator to generate quadrature modulated coherent states. In order to provide immunity against constant thermal fluctuations in space, we use ultra-low $V_\pi$ ($1.7~\text{V}$) phase modulators (EOSPACE) in Sagnac interferometric configuration as the amplitude modulators. This is to avoid the need for using a bias control that necessary for conventional amplitude modulators based on Mach Zehnder interferometers. A portion of the modulated signal is detected with PIN photodiode while the other part is polarisation and time multiplexed with the LO pulse. A homodyne detector for QRNG is constructed with a fibre optic beam splitter and balanced PIN photodiodes. The laser pulse from a portion of the monitoring port serves as the LO for the QRNG. Polarisation maintaining fibre is used through out the optical module. The multiplexed fibre output is collimated, and then polarisation is rotated by a quarter waveplate to the right and left polarisation, before being sent to the ground. The circular polarisations of light gives tolerance to satellite rotation and avoid the necessity of using polarisation correction at the ground.

\begin{widetext}
\begin{center}
\begin{figure}[ht]
\includegraphics[width=0.95\linewidth]{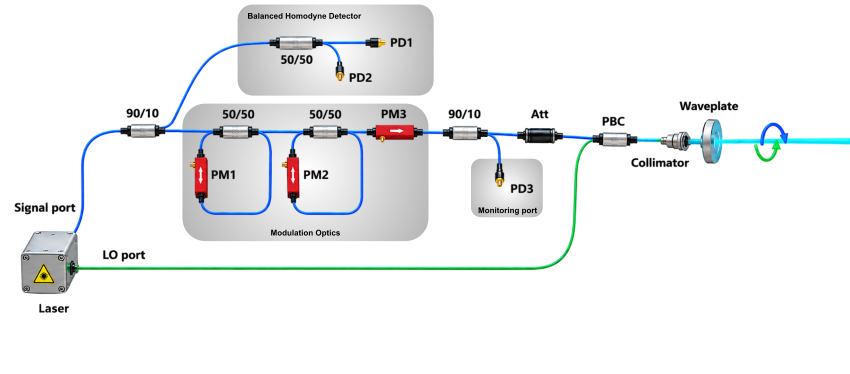}
\caption{The design of the CVP optics is shown here. The $2~\text{ns}$ light at repetition rate of $2~\text{MHz}$ is divided into two ports - local oscillator (green path) and signal port (blue path), with average power in each of them to be of $\sim 1~\text{W}$ and $\sim 2~\text{mW}$, respectively. This produces a circular polarised signal and LO. The "Balanced Homodyne Detector" section represents the QRNG module. Here, X/Y - beam splitters; PM - phase modulator; PD - photodiode; PBC - polarising beam combiner; Att - Attenuator.}
\label{fig:design}
\end{figure}
\end{center}
\end{widetext}

Additionally, since we have the continuous variable-QRNG (CV-QRNG) module in the CVP, which includes a set of InGaAs PIN diodes along with signal amplifier and analog to digital converter, all these have to be space qualified with minimal change in performance. Below we address each unit, with standards of system performance accounted for.

\subsection{Electronics module \label{subsec:ele_mod}}

The electronic module consists of SBC and a daughter board. The daughter board distributes the power to SBC, laser and modulator drivers through a current limiting and shutdown circuitry which provides protection against over current, in the case of single event upset or device failure. It also provides an operational versatility in terms of selectively switching of QRNG, laser or modulations drivers as per the payload concept-of-operation. The daughter board also includes a transimpedance amplifier its input signal can be digitally selected between the monitoring PIN photodiode and QRNG balanced photodiodes. The amplifier output is digitised by a 12-bit analog to digital converter fed to the digital ports of the SBC. The final segment of the daughter board is the drivers and control electronics for the modulators to for the GMCS protocol.

We use Q8S from Xiphos System Corporation as the SBC that works on the Linux operating system. Once the power to the payload is established and boot sequence is completed, the Q8 selects a operational methods as per the configuration file system. For the QKD signal generation, onboard field-programmable gate array (FPGA) delivers the pre-loaded modulation data. Whereas for QRNG operations, the ADC outputs are data logged. The SBC also periodically takes house keeping data and pass it to the central processing computer (CPC) upon request. SBC also performs communication to the ground through the CPC. It also process the QKD data for reconciliation and key generation with the ground.

\subsection{Software \label{subsec:software}}

The CVP is controlled by several bespoke software components, packaged into a custom embedded Linux distribution built using the Yocto Project (version 2.5). This distribution is configured to include only the kernel modules, packages, and software required by the CVP. The Yocto Project enables versioned, reproducible builds of the full CVP software stack, as well as straightforward addition and testing of new software components as they are developed and integrated. The pictorial representation of the software architecture is given in Fig. \ref{fig:soft_arch}.

The CVP functionality is driven by communication from the satellite platform Central Processing Computer (CPC) to the CVP communications server (\emph{cvp-comms}), a Python TCP server application that implements the mission command protocol, oversees the execution of other CVP software components, and negotiates the collection of CVP housekeeping data. Core CVP activities are parametrisable using a configuration file uploaded from a ground station. 

Upon instruction from the communications server, this configuration file is read by an additional Python program (\emph{cvp-alice}) - \emph{cvp-alice} interprets the file to carry out appropriate data collection and processing activities, presenting raw data, intermediates, and final results in a compressed archive for retrieval by the ground station. Data from the electronics module is collected using a C program that operates the CVP laser with configurable power and funnels data from the ADCs into data files via a FIFO queue shared by the FPGA and the Linux system. 

Housekeeping quantities such as system temperatures, current system status, and safety flags are continuously monitored by a stand-alone C program (\emph{cvp-housekeeping}), which presents these quantities to \emph{cvp-comms} when housekeeping data is requested by the satellite platform CPC for mission control and monitoring purposes.

\begin{widetext}
\begin{center}
\begin{figure}[ht]
\includegraphics[width=0.75\textwidth]{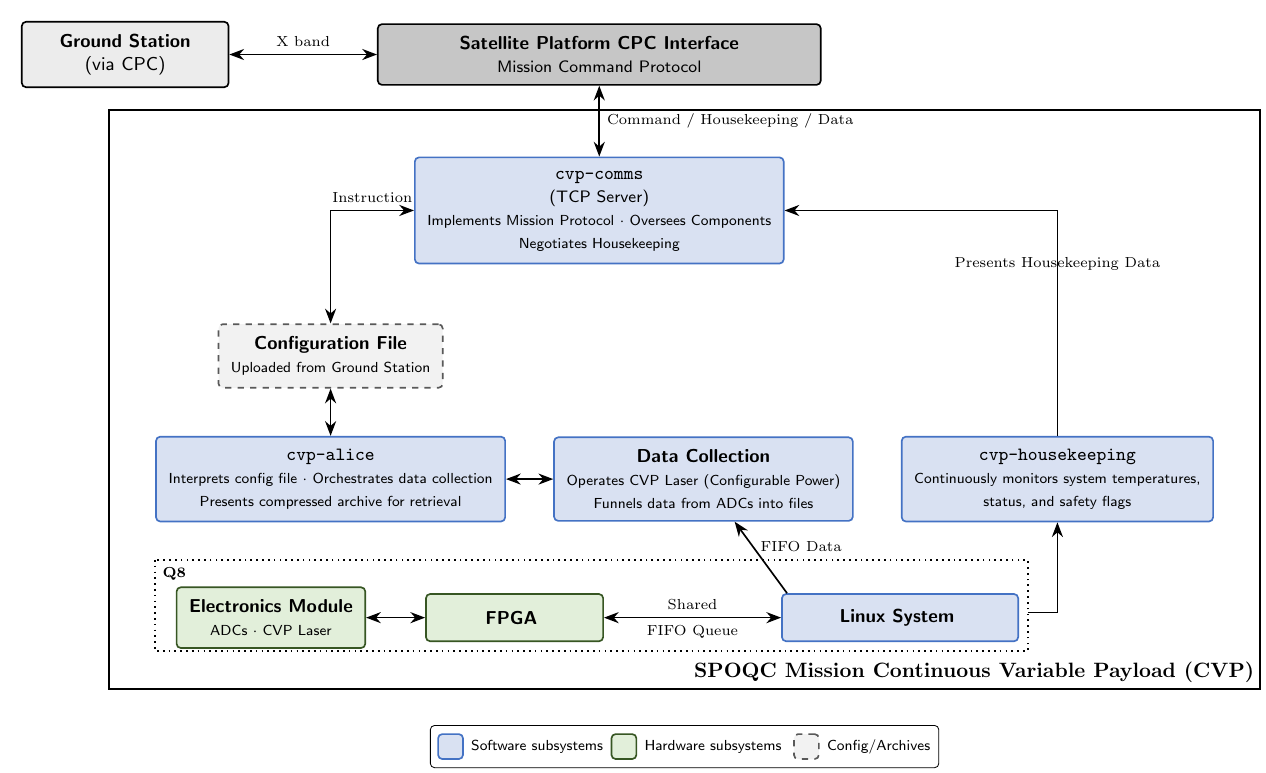}
\caption{The software architecture of the CVP control software is shown. The SBC runs Linux, which hosts the applications \emph{cvp-comms}, \emph{cvp-alice}, and \emph{cvp-housekeeping}.}
\label{fig:soft_arch}
\end{figure}
\end{center}
\end{widetext}

\subsection{Gen of GMCS (QKD) \& random numbers (QRNG) \label{subsec:generation}}

\noindent\textbf{Generation of GMCS:} In CV-QKD, information is encoded in the quadratures of coherent states of the electromagnetic field. In a typical GMCS protocol, the sender (CVP) prepares coherent states whose quadratures are randomly drawn from a centered Gaussian distribution. In this subsection, we provide the details of the GMCS protocol employed in the CVP and show the output to be the same. Here we use the laser with the wavelength output of $1550.12~\text{nm}$ and the pulse width of $2~\text{ns}$. The repetition rate is $2~\text{MHz}$ and the signal \& LO are polarisation multiplexed.

\begin{figure}[h]
\centering
\includegraphics[width=0.95\linewidth]{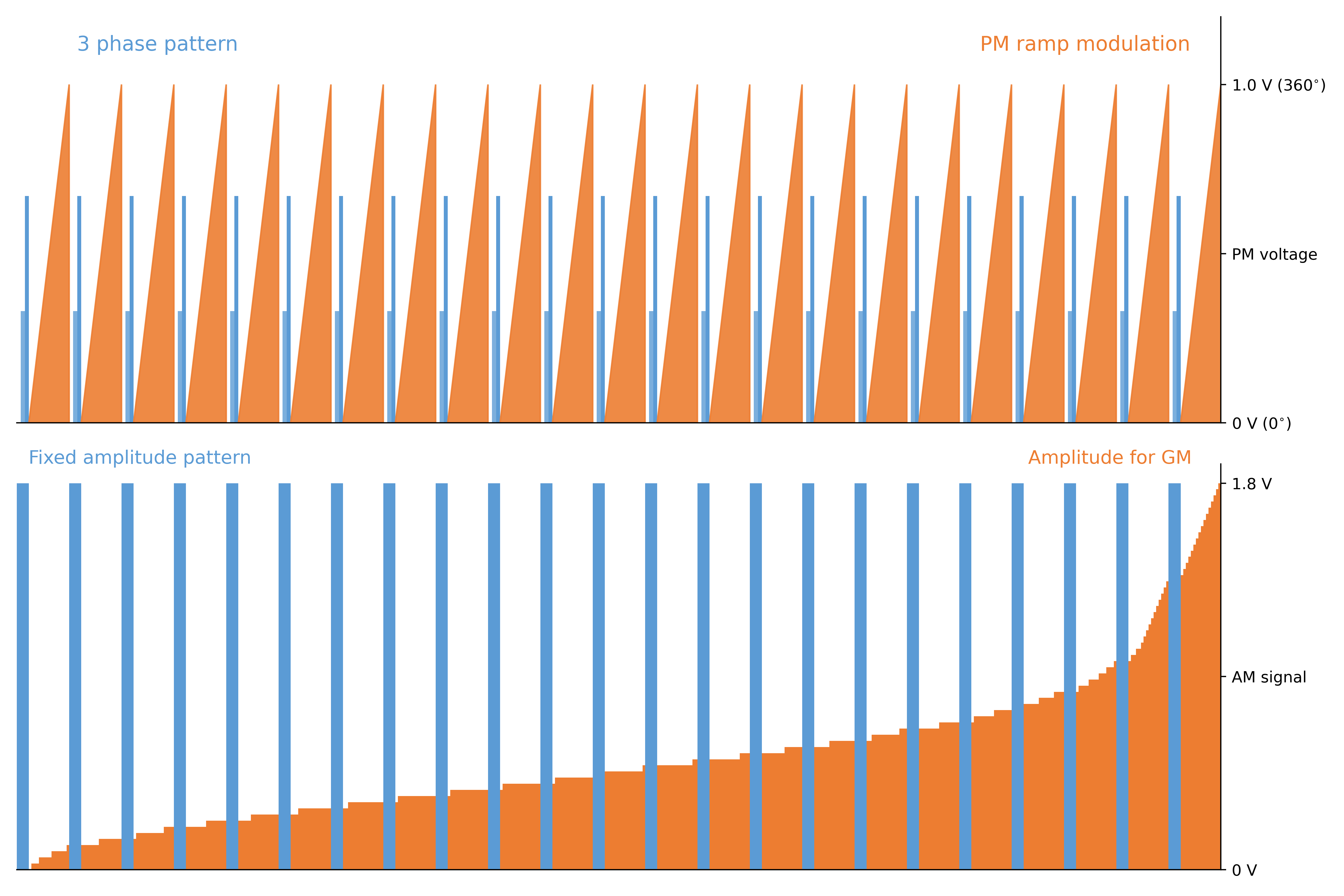}
\caption{The amplitude and phase modulation voltage per data block is shown. Blue is the data for phase pattern and Orange is for QKD signal. The quadrature of the resultant coherent states follows Gaussian distribution. The modulation depth, block size and variance of the data can be changed on demand.}
\label{fig:GM_pattern}
\end{figure}

\begin{figure}[h]
\centering
\includegraphics[width=0.95\linewidth]{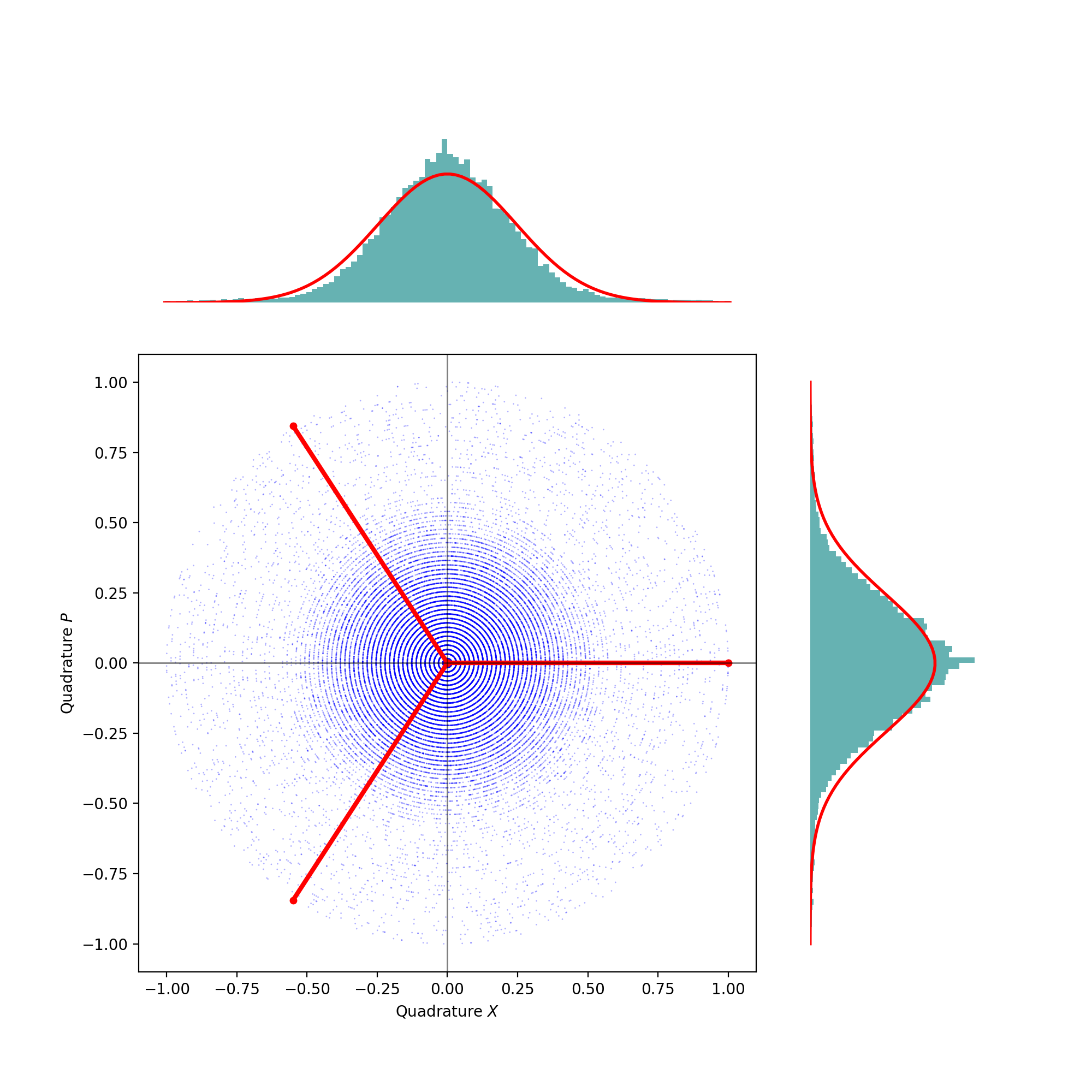}
\caption{The resultant Gaussian modulated coherent states after amplitude modulation and phase modulation. The data here is the input data to CVP. The amplitude of the pattern signal is arbitrarily high (and not scaled here).}
\label{fig:Gaussian_state}
\end{figure}

Through the configuration file, we can select the GMCS modulation variance, operational sequence - QKD or QRNG, and other features of the CVP. In order to generate GMCS, the amplitude and phase modulation data is pre-loaded to the SBC FPGA. The probability distribution of the amplitude data follows Rayliegh distribution while the phase modulation follows sequential rotation from $0^\circ$ to $360^\circ$ in the phase phase. Different modulation variance can be selected with digitally controlled potentiometers in the daughter board. The details of the programming to generate the GMCS in CVP is given below. The amplitude and phase modulation data sequence is as show in Fig. \ref{fig:GM_pattern}. A phase pattern ($32\times 3$ pulses) is inserted periodically within the data blocks ($2048$ pulses) to estimate the relative phase rotation of the signal and LO during the homodyne detection at the OGS. Fixed amplitude is given to these phase patterns ($96$ pulses), with one cycle of amplitude modulation (AM) taking 23 repeated cycles of phase modulation (PM). Based on this design, a single complete cycle of the GMCS requires a total of $96,468,992$ pulses. This total is derived directly from the structural layout of the AM \& PM programming, representing the cumulative product of the signal span and the phase modulation steps as $(23 \times 2,048) \times 2,048 = 96,468,992~\text{pulses}$. The discrete AM voltage levels update precisely once every $128$ pulses. Ultimately, the system maps across $2^{12} = 4,096$ total discrete AM voltage levels and $2^{11} = 2,048$ PM resolution levels, utilizing an underlying structure of $368$ AM data array elements to manage the core configuration across the waveform's entire span. The expected GMCS generated from the AM and PM in CVP is given in Fig. \ref{fig:Gaussian_state}. The downlink classical data that is generated in the considered time window, required for parameter estimation, is about $700~\text{MB}$.

\noindent\textbf{Generation of random numbers:} The operational sequence begins by configuring the homodyne detection unit of the CVP in QRNG mode, wherein a digitally controlled relay switch blocks the monitoring photodetector current associated with the QKD signal from reaching the QRNG electronic board. Prior to each QRNG session, electronic noise quadrature data are acquired to account for temperature induced drift in the electronic noise. Although temperature variations may lead to drift in the laser operating wavelength, they do not significantly affect the shot noise level. Other classical noise contributions, such as those arising from imbalance in the homodyne setup and detector inefficiencies, have been calibrated beforehand in laboratory conditions prior to satellite launch. The homodyne detector exhibits a common-mode rejection ratio greater than $50~\text{dB}$, rendering the influence of laser intensity fluctuations negligible. Thus, the input noise is assumed to be predominantly thermal.

\begin{figure}[h]
\centering
\includegraphics[width=0.95\linewidth]{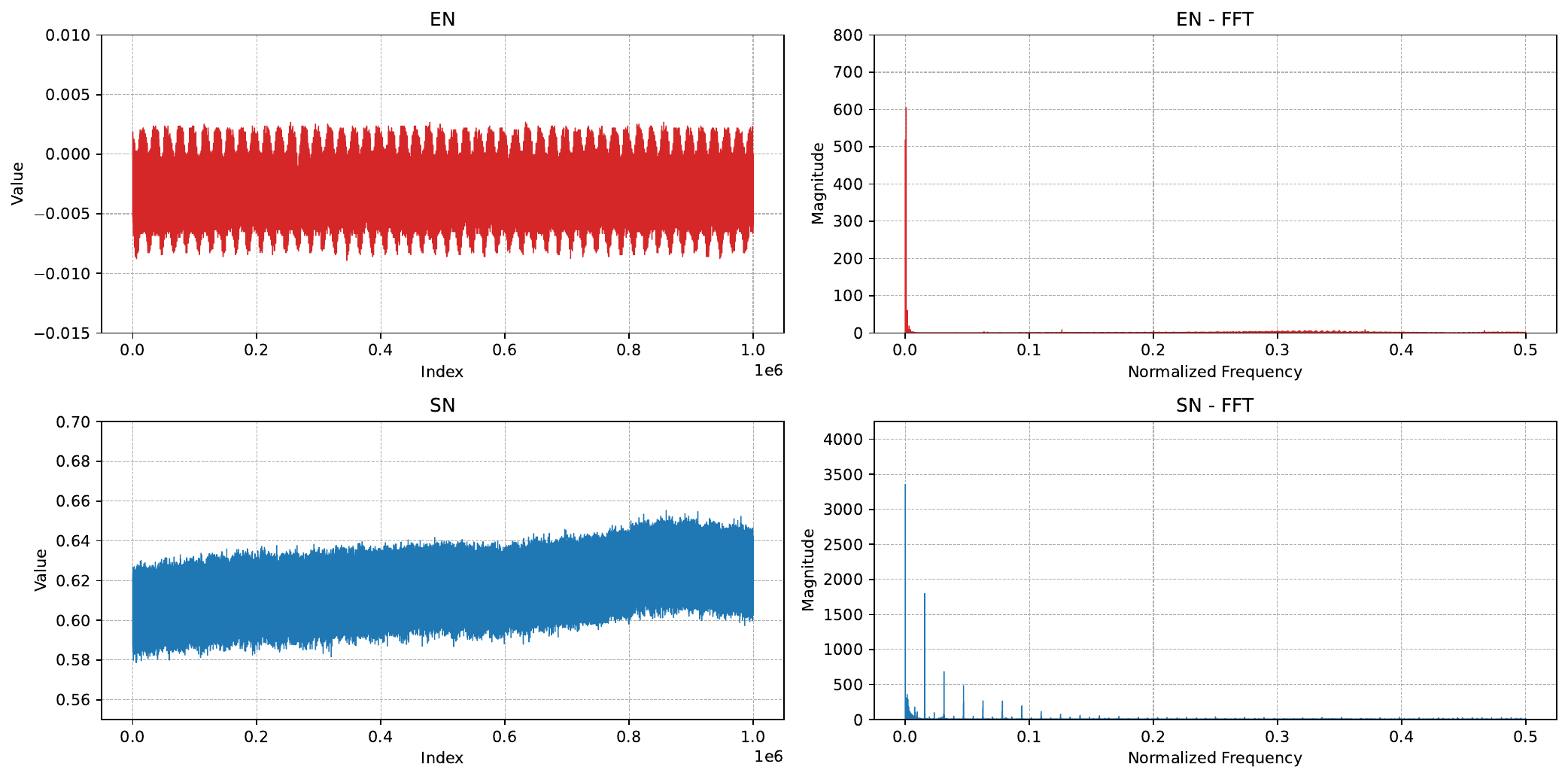}
\caption{A sample set of raw data of the electronic noise and shot noise from the CV-QRNG module in the CVP and Fourier transform of the same.}
\label{fig:Raw_data}
\end{figure}

Subsequently, the laser is switched on and the shot noise quadrature data are acquired. The QRNG optical subsystem comprises a radiation-hardened 50/50 fibre beam splitter and two vacuum compatible PIN photodiodes. The resulting photocurrents are differentially subtracted and amplified using low noise amplifiers (AMPTEK A250F), followed by digitization using a 12-bit ADC (Analog Devices AD9220). The ADC outputs are then acquired by the SBC at a sampling rate of $500~\text{kHz}$, which is limited by the hardware architecture. A sample output of the electronic noise and shot noise are shown in Fig. \ref{fig:Raw_data}.

\section{Verification and validation \label{sec:test}}

In this section, we provide evidence for the functional performance of the CVP design components - generation of GMCS signals and generation of random numbers for the QRNG. The Gaussian modulation has been verified onboard by collecting data from the photodiode for monitoring the modulated signals. The photodiode can only register the amplitude of the generated signals, however, the quadrature distribution of the signals can be inferred by knowing the phase data. We also conducted the homodyne detection of the signals with the LO, that shows (approximated) Gaussian probability distribution as shown in Fig. \ref{fig:CVP_output}. The estimated variance of the distribution for the $30~\text{dB}$ loss channel is found to be $V_A = 85$.

\subsection{Verification of GMCS \label{subsec:verify_GMCS}}

The CVP is designed to generate Gaussian modulated coherent states, with the optical setup given in Fig. \ref{fig:design}. To verify the generation of GMCS by the CVP, a series of measurements were performed on the optical output using the on board modulation and detection electronics. The amplitude and phase modulators were driven according to a pre-defined discretised Gaussian distribution as shown in Fig. \ref{fig:GM_pattern}, resulting in a set of coherent states whose quadrature displacements follow the target modulation profile.

\begin{figure}[ht]
\centering
\begin{subfigure}{0.5\textwidth}
\includegraphics[width=0.95\linewidth]{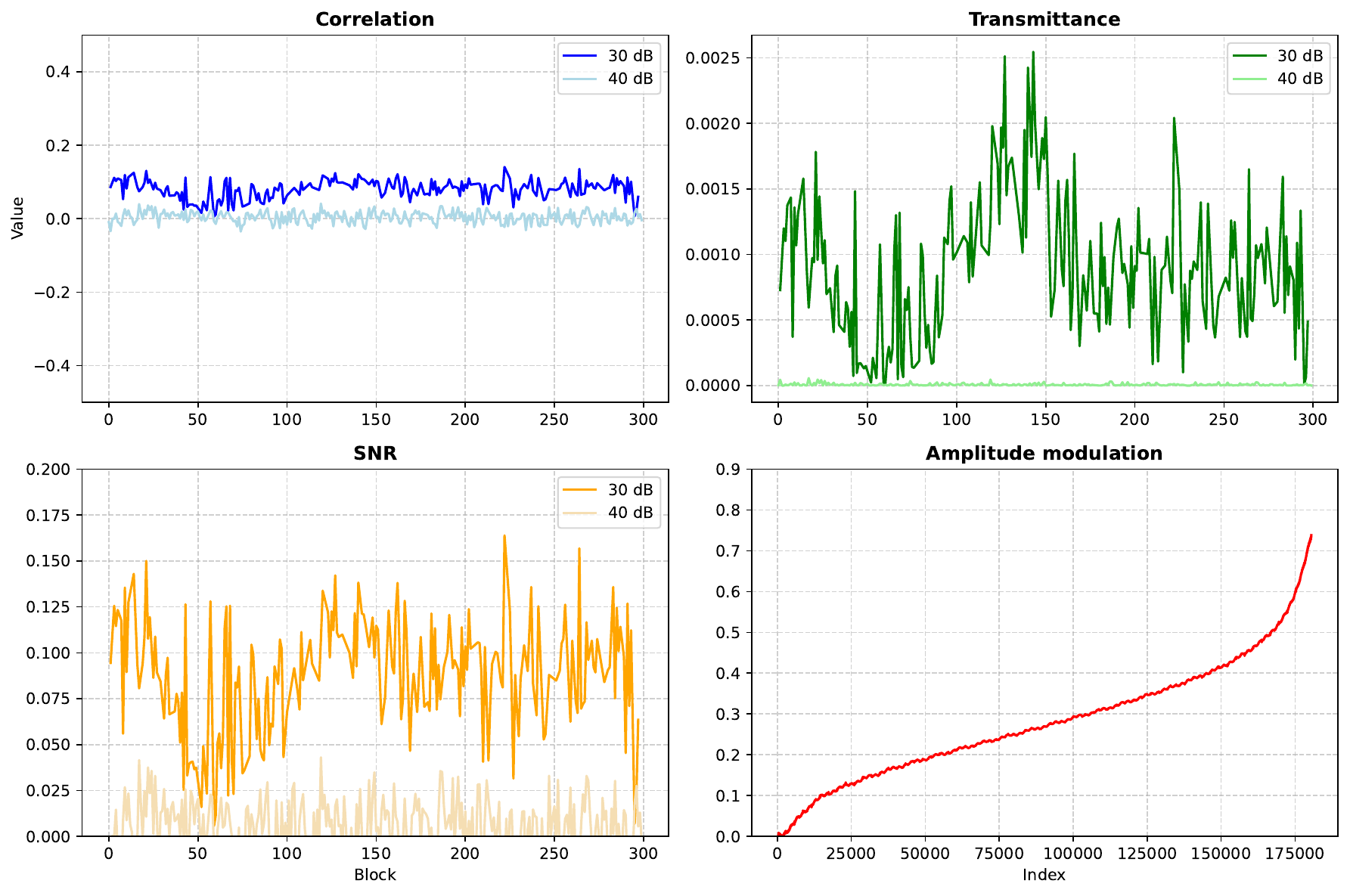}
\caption{}
\label{fig:GM_proof} 
\end{subfigure}
\medskip
\begin{subfigure}{0.5\textwidth}
\includegraphics[width=0.95\linewidth]{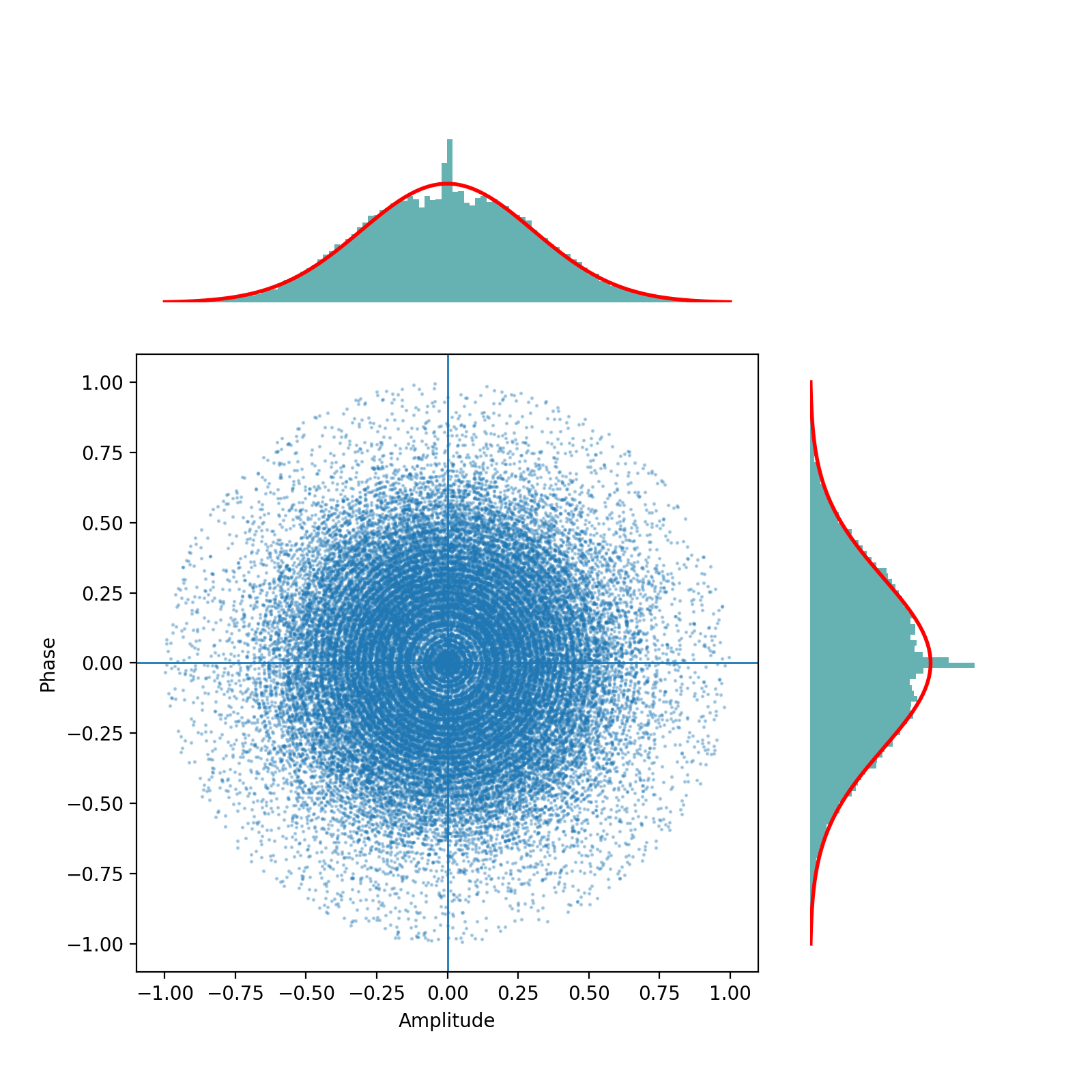}
\caption{}
\label{fig:CVP_GMCS}
\end{subfigure}
\caption{Fig. \ref{fig:GM_proof}: Image showing different parameters measured from the photodiode in the CVP. Each block has a data of $\approx 10^6$ measured values, and the average correlation and SNR are positive. The darker line is for $30~\text{dB}$ lossy channel and the lighter line for $40~\text{dB}$. Fig. \ref{fig:CVP_GMCS}: The filtered and processed output from the CVP showing the generation of GMCS.}
\label{fig:CVP_output}
\end{figure}

Fig. \ref{fig:CVP_output} shows the output of various parameters measured from the data obtained of CVP. The generated coherent states exhibit the expected circularly symmetric Gaussian modulation in the quadrature plane, demonstrating that the modulation electronics and optical subsystem accurately reproduce the intended probability distribution. The corresponding marginal distributions of the $X$ and $P$ quadratures are also shown and are observed to closely follow Gaussian statistics.

The outputs of various parameters obtained from CVP data shows the verification of the working of the CVP, as in Fig. \ref{fig:CVP_output}. From Fig. \ref{fig:GM_proof}, the average correlation and average SNR values for a $30~\text{dB}$ ($40~\text{dB}$) loss channel are found to be $0.08054$ ($0.00413$) and $0.08946$ ($0.0044$), respectively. In the first three sub-figures, each block has a data of $\approx 10^6$ measured values. The last sub-figure in Fig. \ref{fig:GM_proof} shows the amplitude modulation in the CVP, from the photodiode data in the monitoring port. The fluctuation in the estimated parameters in Fig. \ref{fig:GM_proof} is due to the fact that the photodiode averages the output due to its lower sampling rate.

Fig. \ref{fig:CVP_GMCS} shows the generation of GMCS in the CVP. The output with the normalised axis values has the same number of data points as in Fig. \ref{fig:Gaussian_state} (47,104). However, the slight deviation from the Gaussianity is due to the presence of values being sampled more than others. Additionally, the presence of more vacuum states can also be seen as the drop in the distribution near origin.

The verification was repeated after environmental qualification testing, including thermal cycling and vacuum exposure, to ensure that the modulation characteristics remained stable under space relevant conditions. No statistically significant changes were observed in the measured distributions, confirming the robustness of the CVP optical and electronic subsystems. These results demonstrate the successful generation of GMCS by the CVP and validate its suitability for demonstration of downlink based CV-QKD (and CV-QRNG).

\subsection{Verification of QRNG \label{subsec:verify_QRNG}}

The QRNG module in the CVP contains a homodyne detector followed by a 12-bit ADC. Additionally, SBC is also present where the randomness extraction happens. In Fig. \ref{fig:qrng}, the output from the NIST test suite of the CV-QRNG.

We account for quantum side-information leakage to a potential eavesdropper Eve by deriving an upper bound on the conditional min-entropy $H_{\text{min}}$. In addition, noise arising from the non-ideal behaviour of the ADC is incorporated into the analysis. Following the estimation of the conditional min-entropy, randomness extraction is performed using Toeplitz matrix hashing. We choose a hashing failure probability of $\epsilon_{\text{hash}} = 10^{-20}$, which ensures that the overall security parameter remains within the conventional bound of $10^{-10}$. \cite{pirandola2020advances}. The Toeplitz matrix is configured with dimensions $900 \times 200$, yielding 200 bits of extracted randomness for every 100 ADC samples (corresponding to $0.1~\text{ms}$). Given a raw key length of approximately 1 Mb, this results in about 19.5 Kb of certified random numbers from the 12-bit ADC. Results for a 16-bit ADC are also presented, although these are not obtained from the on-board system. A key aspect for future investigation is the impact of temperature variations on system components and the corresponding optimization of the conditional min-entropy. Further details of this work can be found in Ref. \cite{nagaraja2026quantum}.

\subsection{Environmental testing \label{subsec:env_test}}

The CVP was tested at various stages to confirm its working. While the initial in house testing included purely functional tests of the CVP using a direct connection to a test computer, we provide the details of the testing from the last three stages - environmental tests, test for showing GM after environmental tests and a brief comment on testing after the integration to the CubeSat. And at each stage, both the optics module and electronics module was tested. Additionally, we also provide details of the outcome from the QRNG module. The details of the QRNG module is given in appendix Sec. \ref{subsec:opt_mod}.

The environmental qualification tests involved subjecting the CVP to a range of conditions, including temperature, pressure, shock, and vibration testing. These tests were conducted at RAL Space, where the CVP module was evaluated over a temperature range of $-20~^{\circ}\text{C}$ to $+55~^{\circ}\text{C}$ and under pressure conditions varying from $1~\text{atm}$ down to $10^{-6}~\text{atm}$. Additionally, shock testing was performed up to 10 G, while vibration testing was carried out over a frequency range of $0-200~\text{Hz}$. We note that during the TVAC (temperature \& vacuum/pressure) testing, the output from the CVP was constantly monitored. Following the environmental testing, the CVP was re-evaluated in our laboratory to verify both its performance and functionality. Subsequently, final functional tests were carried out after the integration of the CVP with the CubeSat platform.

\section{Discussions and Conclusions \label{sec:disc}}

In this work, we have presented the design, implementation, and experimental validation of the CV quantum payload for a downlink based quantum communication from a CubeSat. The system integrates both the optical and electronic subsystems required for CV-QKD, while adhering to stringent size, weight, and power (SWaP) constraints inherent to space based platforms. The successful realization of a compact and robust architecture demonstrates the feasibility of deploying CV quantum technologies on resource limited satellite platforms. The experimental characterization confirms the reliable operation of the payload, including on board stable homodyne detection for CV-QRNG, controlled local oscillator distribution, and consistent performance of the electronic systems. Furthermore, the incorporation of realistic noise sources, such as electronic noise and detector imperfections, enables a conservative estimation of the extractable secure randomness for the CV-QRNG. 

A key outcome of this work is the demonstration that CV-QKD protocols can be adapted to operate under the practical limitations of a CubeSat environment. In particular, constraints on data acquisition rates, LO power, and thermal stability have been carefully addressed. The results indicate that, despite these limitations, meaningful secure key rates can be achieved, thereby highlighting the potential of CV-based approaches for satellite quantum communications. The future CV payloads can be optimised by using small form factor components, migrating to LLO based scheme, or using photonic integrated chips bringing the size to $\le 0.5~\text{U}$, weight to a few $100\text{s}$ of grams and power down to $\le 5~\text{W}$ \cite{vokes2026energy}. 

Future work will focus on in orbit validation of the payload, including the assessment of channel induced noise, real time parameter estimation, and adaptive optimization of the protocol under varying environmental conditions. Additionally, further improvements in entropy estimation, randomness extraction, and thermal management are expected to enhance the overall system performance and security. Finally, we note that a CV receiver (CVR) was specifically developed for the CVP. The design of the CVR is unique to CVP and separate work on that is being developed with the single quadrature measurement \cite{rao2025continuous}. And the requirement of SWaP, along with space qualified instruments, is relaxed in the CVR.

In conclusion, this work establishes a practical pathway for implementing CV-QKD on CubeSat platforms and represents a significant step toward scalable, space based quantum communication networks. The demonstrated system lays the groundwork for future missions aimed at global quantum key distribution and secure satellite to ground links.

\begin{acknowledgments}
The authors acknowledge the funding support from EPSRC Quantum Communications Hub (Grant number EP/T001011/1) and EPSRC Integrated Quantum Networks Hub (Grant No. EP/Z533208/1). Additionally, the authors also thank RAL Space for conducting the Environmental test on the CV payload.
\end{acknowledgments}

The data that support the findings of this study are openly available at the following URL/DOI: \href{https://doi.org/10.15124/4354451f-30e0-40d1-8073-1fe7a1406681}{\textcolor{blue}{https://doi.org/10.15124/4354451f-30e0-40d1-8073-1fe7a1406681}}.

\bibliography{References}

\begin{thebibliography}{32}%
\makeatletter
\providecommand \@ifxundefined [1]{%
 \@ifx{#1\undefined}
}%
\providecommand \@ifnum [1]{%
 \ifnum #1\expandafter \@firstoftwo
 \else \expandafter \@secondoftwo
 \fi
}%
\providecommand \@ifx [1]{%
 \ifx #1\expandafter \@firstoftwo
 \else \expandafter \@secondoftwo
 \fi
}%
\providecommand \natexlab [1]{#1}%
\providecommand \enquote  [1]{``#1''}%
\providecommand \bibnamefont  [1]{#1}%
\providecommand \bibfnamefont [1]{#1}%
\providecommand \citenamefont [1]{#1}%
\providecommand \href@noop [0]{\@secondoftwo}%
\providecommand \href [0]{\begingroup \@sanitize@url \@href}%
\providecommand \@href[1]{\@@startlink{#1}\@@href}%
\providecommand \@@href[1]{\endgroup#1\@@endlink}%
\providecommand \@sanitize@url [0]{\catcode `\\12\catcode `\$12\catcode `\&12\catcode `\#12\catcode `\^12\catcode `\_12\catcode `\%12\relax}%
\providecommand \@@startlink[1]{}%
\providecommand \@@endlink[0]{}%
\providecommand \url  [0]{\begingroup\@sanitize@url \@url }%
\providecommand \@url [1]{\endgroup\@href {#1}{\urlprefix }}%
\providecommand \urlprefix  [0]{URL }%
\providecommand \Eprint [0]{\href }%
\providecommand \doibase [0]{https://doi.org/}%
\providecommand \selectlanguage [0]{\@gobble}%
\providecommand \bibinfo  [0]{\@secondoftwo}%
\providecommand \bibfield  [0]{\@secondoftwo}%
\providecommand \translation [1]{[#1]}%
\providecommand \BibitemOpen [0]{}%
\providecommand \bibitemStop [0]{}%
\providecommand \bibitemNoStop [0]{.\EOS\space}%
\providecommand \EOS [0]{\spacefactor3000\relax}%
\providecommand \BibitemShut  [1]{\csname bibitem#1\endcsname}%
\let\auto@bib@innerbib\@empty
\bibitem [{\citenamefont {Bennett}\ and\ \citenamefont {Brassard}(1984)}]{bennett1984quantum}%
  \BibitemOpen
  \bibfield  {author} {\bibinfo {author} {\bibfnamefont {C.~H.}\ \bibnamefont {Bennett}}\ and\ \bibinfo {author} {\bibfnamefont {G.}~\bibnamefont {Brassard}},\ }\bibfield  {title} {\bibinfo {title} {Quantum cryptography: Public key distribution and coin tossing},\ }in\ \href@noop {} {\emph {\bibinfo {booktitle} {Proceedings of the International Conference on Computers, Systems and Signal Processing}}}\ (\bibinfo {year} {1984})\BibitemShut {NoStop}%
\bibitem [{\citenamefont {Ekert}(1991)}]{ekert1991quantum}%
  \BibitemOpen
  \bibfield  {author} {\bibinfo {author} {\bibfnamefont {A.~K.}\ \bibnamefont {Ekert}},\ }\bibfield  {title} {\bibinfo {title} {Quantum cryptography based on bell’s theorem},\ }\href@noop {} {\bibfield  {journal} {\bibinfo  {journal} {Physical Review Letters}\ }\textbf {\bibinfo {volume} {67}},\ \bibinfo {pages} {661} (\bibinfo {year} {1991})}\BibitemShut {NoStop}%
\bibitem [{\citenamefont {Bennett}\ \emph {et~al.}(1992)\citenamefont {Bennett}, \citenamefont {Brassard},\ and\ \citenamefont {Mermin}}]{bennett1992quantum2}%
  \BibitemOpen
  \bibfield  {author} {\bibinfo {author} {\bibfnamefont {C.~H.}\ \bibnamefont {Bennett}}, \bibinfo {author} {\bibfnamefont {G.}~\bibnamefont {Brassard}},\ and\ \bibinfo {author} {\bibfnamefont {N.~D.}\ \bibnamefont {Mermin}},\ }\bibfield  {title} {\bibinfo {title} {Quantum cryptography without bell’s theorem},\ }\href@noop {} {\bibfield  {journal} {\bibinfo  {journal} {Physical Review Letters}\ }\textbf {\bibinfo {volume} {68}},\ \bibinfo {pages} {557} (\bibinfo {year} {1992})}\BibitemShut {NoStop}%
\bibitem [{\citenamefont {Lu}\ \emph {et~al.}(2022)\citenamefont {Lu}, \citenamefont {Cao}, \citenamefont {Peng},\ and\ \citenamefont {Pan}}]{lu2022micius}%
  \BibitemOpen
  \bibfield  {author} {\bibinfo {author} {\bibfnamefont {C.-Y.}\ \bibnamefont {Lu}}, \bibinfo {author} {\bibfnamefont {Y.}~\bibnamefont {Cao}}, \bibinfo {author} {\bibfnamefont {C.-Z.}\ \bibnamefont {Peng}},\ and\ \bibinfo {author} {\bibfnamefont {J.-W.}\ \bibnamefont {Pan}},\ }\bibfield  {title} {\bibinfo {title} {Micius quantum experiments in space},\ }\href@noop {} {\bibfield  {journal} {\bibinfo  {journal} {Reviews of Modern Physics}\ }\textbf {\bibinfo {volume} {94}},\ \bibinfo {pages} {035001} (\bibinfo {year} {2022})}\BibitemShut {NoStop}%
\bibitem [{\citenamefont {Li}\ \emph {et~al.}(2025)\citenamefont {Li}, \citenamefont {Cai}, \citenamefont {Ren}, \citenamefont {Wang}, \citenamefont {Yang}, \citenamefont {Zhang}, \citenamefont {Wu}, \citenamefont {Chang}, \citenamefont {Wu}, \citenamefont {Jin} \emph {et~al.}}]{li2025microsatellite}%
  \BibitemOpen
  \bibfield  {author} {\bibinfo {author} {\bibfnamefont {Y.}~\bibnamefont {Li}}, \bibinfo {author} {\bibfnamefont {W.-Q.}\ \bibnamefont {Cai}}, \bibinfo {author} {\bibfnamefont {J.-G.}\ \bibnamefont {Ren}}, \bibinfo {author} {\bibfnamefont {C.-Z.}\ \bibnamefont {Wang}}, \bibinfo {author} {\bibfnamefont {M.}~\bibnamefont {Yang}}, \bibinfo {author} {\bibfnamefont {L.}~\bibnamefont {Zhang}}, \bibinfo {author} {\bibfnamefont {H.-Y.}\ \bibnamefont {Wu}}, \bibinfo {author} {\bibfnamefont {L.}~\bibnamefont {Chang}}, \bibinfo {author} {\bibfnamefont {J.-C.}\ \bibnamefont {Wu}}, \bibinfo {author} {\bibfnamefont {B.}~\bibnamefont {Jin}}, \emph {et~al.},\ }\bibfield  {title} {\bibinfo {title} {Microsatellite-based real-time quantum key distribution},\ }\href@noop {} {\bibfield  {journal} {\bibinfo  {journal} {Nature}\ ,\ \bibinfo {pages} {1}} (\bibinfo {year} {2025})}\BibitemShut {NoStop}%
\bibitem [{\citenamefont {Takenaka}\ \emph {et~al.}(2017)\citenamefont {Takenaka}, \citenamefont {Carrasco-Casado}, \citenamefont {Fujiwara}, \citenamefont {Kitamura}, \citenamefont {Sasaki},\ and\ \citenamefont {Toyoshima}}]{takenaka2017satellite}%
  \BibitemOpen
  \bibfield  {author} {\bibinfo {author} {\bibfnamefont {H.}~\bibnamefont {Takenaka}}, \bibinfo {author} {\bibfnamefont {A.}~\bibnamefont {Carrasco-Casado}}, \bibinfo {author} {\bibfnamefont {M.}~\bibnamefont {Fujiwara}}, \bibinfo {author} {\bibfnamefont {M.}~\bibnamefont {Kitamura}}, \bibinfo {author} {\bibfnamefont {M.}~\bibnamefont {Sasaki}},\ and\ \bibinfo {author} {\bibfnamefont {M.}~\bibnamefont {Toyoshima}},\ }\bibfield  {title} {\bibinfo {title} {Satellite-to-ground quantum-limited communication using a 50-kg-class microsatellite},\ }\href@noop {} {\bibfield  {journal} {\bibinfo  {journal} {Nature photonics}\ }\textbf {\bibinfo {volume} {11}},\ \bibinfo {pages} {502} (\bibinfo {year} {2017})}\BibitemShut {NoStop}%
\bibitem [{\citenamefont {Villar}\ \emph {et~al.}(2020)\citenamefont {Villar}, \citenamefont {Lohrmann}, \citenamefont {Bai}, \citenamefont {Vergoossen}, \citenamefont {Bedington}, \citenamefont {Perumangatt}, \citenamefont {Lim}, \citenamefont {Islam}, \citenamefont {Reezwana}, \citenamefont {Tang} \emph {et~al.}}]{villar2020entanglement}%
  \BibitemOpen
  \bibfield  {author} {\bibinfo {author} {\bibfnamefont {A.}~\bibnamefont {Villar}}, \bibinfo {author} {\bibfnamefont {A.}~\bibnamefont {Lohrmann}}, \bibinfo {author} {\bibfnamefont {X.}~\bibnamefont {Bai}}, \bibinfo {author} {\bibfnamefont {T.}~\bibnamefont {Vergoossen}}, \bibinfo {author} {\bibfnamefont {R.}~\bibnamefont {Bedington}}, \bibinfo {author} {\bibfnamefont {C.}~\bibnamefont {Perumangatt}}, \bibinfo {author} {\bibfnamefont {H.~Y.}\ \bibnamefont {Lim}}, \bibinfo {author} {\bibfnamefont {T.}~\bibnamefont {Islam}}, \bibinfo {author} {\bibfnamefont {A.}~\bibnamefont {Reezwana}}, \bibinfo {author} {\bibfnamefont {Z.}~\bibnamefont {Tang}}, \emph {et~al.},\ }\bibfield  {title} {\bibinfo {title} {Entanglement demonstration on board a nano-satellite},\ }\href@noop {} {\bibfield  {journal} {\bibinfo  {journal} {Optica}\ }\textbf {\bibinfo {volume} {7}},\ \bibinfo {pages} {734} (\bibinfo {year} {2020})}\BibitemShut {NoStop}%
\bibitem [{\citenamefont {Knips}\ \emph {et~al.}(2022)\citenamefont {Knips}, \citenamefont {Auer}, \citenamefont {Baliuka}, \citenamefont {Bayraktar}, \citenamefont {Freiwang}, \citenamefont {Gr{\"u}nefeld}, \citenamefont {Haber}, \citenamefont {Lemke}, \citenamefont {Marquardt}, \citenamefont {Moll} \emph {et~al.}}]{knips2022qube}%
  \BibitemOpen
  \bibfield  {author} {\bibinfo {author} {\bibfnamefont {L.}~\bibnamefont {Knips}}, \bibinfo {author} {\bibfnamefont {M.}~\bibnamefont {Auer}}, \bibinfo {author} {\bibfnamefont {A.}~\bibnamefont {Baliuka}}, \bibinfo {author} {\bibfnamefont {{\"O}.}~\bibnamefont {Bayraktar}}, \bibinfo {author} {\bibfnamefont {P.}~\bibnamefont {Freiwang}}, \bibinfo {author} {\bibfnamefont {M.}~\bibnamefont {Gr{\"u}nefeld}}, \bibinfo {author} {\bibfnamefont {R.}~\bibnamefont {Haber}}, \bibinfo {author} {\bibfnamefont {N.}~\bibnamefont {Lemke}}, \bibinfo {author} {\bibfnamefont {C.}~\bibnamefont {Marquardt}}, \bibinfo {author} {\bibfnamefont {F.}~\bibnamefont {Moll}}, \emph {et~al.},\ }\bibfield  {title} {\bibinfo {title} {Qube--towards quantum key distribution with small satellites},\ }in\ \href@noop {} {\emph {\bibinfo {booktitle} {Quantum 2.0}}}\ (\bibinfo {organization} {Optica Publishing Group},\ \bibinfo {year} {2022})\ pp.\ \bibinfo {pages} {QTh3A--6}\BibitemShut {NoStop}%
\bibitem [{Spe()}]{Speqtre}%
  \BibitemOpen
  \href@noop {} {\bibinfo {title} {Speqtre cubesat mission}},\ \bibinfo {howpublished} {\url{https://speqtralquantum.com/technology/satellite-qkd}}\BibitemShut {NoStop}%
\bibitem [{\citenamefont {Scott}\ \emph {et~al.}(2020)\citenamefont {Scott}, \citenamefont {Jennewein}, \citenamefont {Cain}, \citenamefont {D'Souza}, \citenamefont {Higgins}, \citenamefont {Hudson}, \citenamefont {Podmore},\ and\ \citenamefont {Soh}}]{scott2020qeyssat}%
  \BibitemOpen
  \bibfield  {author} {\bibinfo {author} {\bibfnamefont {A.}~\bibnamefont {Scott}}, \bibinfo {author} {\bibfnamefont {T.}~\bibnamefont {Jennewein}}, \bibinfo {author} {\bibfnamefont {J.}~\bibnamefont {Cain}}, \bibinfo {author} {\bibfnamefont {I.}~\bibnamefont {D'Souza}}, \bibinfo {author} {\bibfnamefont {B.}~\bibnamefont {Higgins}}, \bibinfo {author} {\bibfnamefont {D.}~\bibnamefont {Hudson}}, \bibinfo {author} {\bibfnamefont {H.}~\bibnamefont {Podmore}},\ and\ \bibinfo {author} {\bibfnamefont {W.}~\bibnamefont {Soh}},\ }\bibfield  {title} {\bibinfo {title} {The qeyssat mission: on-orbit demonstration of secure optical communications network technologies},\ }in\ \href@noop {} {\emph {\bibinfo {booktitle} {Environmental effects on light propagation and adaptive systems III}}},\ Vol.\ \bibinfo {volume} {11532}\ (\bibinfo {organization} {SPIE},\ \bibinfo {year} {2020})\ pp.\ \bibinfo {pages} {71--76}\BibitemShut {NoStop}%
\bibitem [{\citenamefont {Armengol}\ \emph {et~al.}(2008)\citenamefont {Armengol}, \citenamefont {Furch}, \citenamefont {de~Matos}, \citenamefont {Minster}, \citenamefont {Cacciapuoti}, \citenamefont {Pfennigbauer}, \citenamefont {Aspelmeyer}, \citenamefont {Jennewein}, \citenamefont {Ursin}, \citenamefont {Schmitt-Manderbach} \emph {et~al.}}]{armengol2008quantum}%
  \BibitemOpen
  \bibfield  {author} {\bibinfo {author} {\bibfnamefont {J.~M.~P.}\ \bibnamefont {Armengol}}, \bibinfo {author} {\bibfnamefont {B.}~\bibnamefont {Furch}}, \bibinfo {author} {\bibfnamefont {C.~J.}\ \bibnamefont {de~Matos}}, \bibinfo {author} {\bibfnamefont {O.}~\bibnamefont {Minster}}, \bibinfo {author} {\bibfnamefont {L.}~\bibnamefont {Cacciapuoti}}, \bibinfo {author} {\bibfnamefont {M.}~\bibnamefont {Pfennigbauer}}, \bibinfo {author} {\bibfnamefont {M.}~\bibnamefont {Aspelmeyer}}, \bibinfo {author} {\bibfnamefont {T.}~\bibnamefont {Jennewein}}, \bibinfo {author} {\bibfnamefont {R.}~\bibnamefont {Ursin}}, \bibinfo {author} {\bibfnamefont {T.}~\bibnamefont {Schmitt-Manderbach}}, \emph {et~al.},\ }\bibfield  {title} {\bibinfo {title} {Quantum communications at esa: Towards a space experiment on the iss},\ }\href@noop {} {\bibfield  {journal} {\bibinfo  {journal} {Acta Astronautica}\ }\textbf {\bibinfo {volume} {63}},\ \bibinfo {pages} {165} (\bibinfo {year} {2008})}\BibitemShut {NoStop}%
\bibitem [{\citenamefont {Jennewein}\ \emph {et~al.}(2014)\citenamefont {Jennewein}, \citenamefont {Grant}, \citenamefont {Choi}, \citenamefont {Pugh}, \citenamefont {Holloway}, \citenamefont {Bourgoin}, \citenamefont {Hakima}, \citenamefont {Higgins},\ and\ \citenamefont {Zee}}]{jennewein2014nanoqey}%
  \BibitemOpen
  \bibfield  {author} {\bibinfo {author} {\bibfnamefont {T.}~\bibnamefont {Jennewein}}, \bibinfo {author} {\bibfnamefont {C.}~\bibnamefont {Grant}}, \bibinfo {author} {\bibfnamefont {E.}~\bibnamefont {Choi}}, \bibinfo {author} {\bibfnamefont {C.}~\bibnamefont {Pugh}}, \bibinfo {author} {\bibfnamefont {C.}~\bibnamefont {Holloway}}, \bibinfo {author} {\bibfnamefont {J.}~\bibnamefont {Bourgoin}}, \bibinfo {author} {\bibfnamefont {H.}~\bibnamefont {Hakima}}, \bibinfo {author} {\bibfnamefont {B.}~\bibnamefont {Higgins}},\ and\ \bibinfo {author} {\bibfnamefont {R.}~\bibnamefont {Zee}},\ }\bibfield  {title} {\bibinfo {title} {The nanoqey mission: ground to space quantum key and entanglement distribution using a nanosatellite},\ }in\ \href@noop {} {\emph {\bibinfo {booktitle} {Emerging technologies in security and defence II; and quantum-physics-based information security III}}},\ Vol.\ \bibinfo {volume} {9254}\ (\bibinfo {organization} {SPIE},\ \bibinfo {year} {2014})\ p.\ \bibinfo {pages} {925402}\BibitemShut
  {NoStop}%
\bibitem [{\citenamefont {Balakier}\ \emph {et~al.}(2025)\citenamefont {Balakier}, \citenamefont {Acar}, \citenamefont {Dequal}, \citenamefont {Scarinzi}, \citenamefont {Calvo}, \citenamefont {Vasko}, \citenamefont {Perdigues},\ and\ \citenamefont {Hauschildt}}]{balakier2025high}%
  \BibitemOpen
  \bibfield  {author} {\bibinfo {author} {\bibfnamefont {K.}~\bibnamefont {Balakier}}, \bibinfo {author} {\bibfnamefont {G.}~\bibnamefont {Acar}}, \bibinfo {author} {\bibfnamefont {D.}~\bibnamefont {Dequal}}, \bibinfo {author} {\bibfnamefont {T.}~\bibnamefont {Scarinzi}}, \bibinfo {author} {\bibfnamefont {R.~M.}\ \bibnamefont {Calvo}}, \bibinfo {author} {\bibfnamefont {C.}~\bibnamefont {Vasko}}, \bibinfo {author} {\bibfnamefont {J.}~\bibnamefont {Perdigues}},\ and\ \bibinfo {author} {\bibfnamefont {H.}~\bibnamefont {Hauschildt}},\ }\bibfield  {title} {\bibinfo {title} {From high throughput optical network to quantum information network: overview of european space agency networks},\ }\href@noop {} {\bibfield  {journal} {\bibinfo  {journal} {Free-Space Laser Communications XXXVII}\ }\textbf {\bibinfo {volume} {13355}},\ \bibinfo {pages} {80} (\bibinfo {year} {2025})}\BibitemShut {NoStop}%
\bibitem [{\citenamefont {Hiemstra}\ \emph {et~al.}(2025)\citenamefont {Hiemstra}, \citenamefont {Hasler}, \citenamefont {Paone}, \citenamefont {Reichert}, \citenamefont {Heine},\ and\ \citenamefont {Struck}}]{hiemstra2025european}%
  \BibitemOpen
  \bibfield  {author} {\bibinfo {author} {\bibfnamefont {T.}~\bibnamefont {Hiemstra}}, \bibinfo {author} {\bibfnamefont {D.}~\bibnamefont {Hasler}}, \bibinfo {author} {\bibfnamefont {D.}~\bibnamefont {Paone}}, \bibinfo {author} {\bibfnamefont {F.}~\bibnamefont {Reichert}}, \bibinfo {author} {\bibfnamefont {F.}~\bibnamefont {Heine}},\ and\ \bibinfo {author} {\bibfnamefont {J.}~\bibnamefont {Struck}},\ }\bibfield  {title} {\bibinfo {title} {The european satellite-based qkd system eagle-1},\ }in\ \href@noop {} {\emph {\bibinfo {booktitle} {Free-space laser communications XXXVII}}},\ Vol.\ \bibinfo {volume} {13355}\ (\bibinfo {organization} {SPIE},\ \bibinfo {year} {2025})\ pp.\ \bibinfo {pages} {216--222}\BibitemShut {NoStop}%
\bibitem [{SPO()}]{SPOQC}%
  \BibitemOpen
  \href@noop {} {\bibinfo {title} {{T}he {H}ub's {S}{P}{O}{Q}{C} mission - {Q}uantum {C}ommunications {H}ub - quantumcommshub.net}},\ \bibinfo {howpublished} {\url{https://www.quantumcommshub.net/research-community/about-the-hub/phase-2/work-package-5/the-hubs-spoqc-mission/}}\BibitemShut {NoStop}%
\bibitem [{\citenamefont {Scarani}\ \emph {et~al.}(2009)\citenamefont {Scarani}, \citenamefont {Bechmann-Pasquinucci}, \citenamefont {Cerf}, \citenamefont {Du{\v{s}}ek}, \citenamefont {L{\"u}tkenhaus},\ and\ \citenamefont {Peev}}]{scarani2009security}%
  \BibitemOpen
  \bibfield  {author} {\bibinfo {author} {\bibfnamefont {V.}~\bibnamefont {Scarani}}, \bibinfo {author} {\bibfnamefont {H.}~\bibnamefont {Bechmann-Pasquinucci}}, \bibinfo {author} {\bibfnamefont {N.~J.}\ \bibnamefont {Cerf}}, \bibinfo {author} {\bibfnamefont {M.}~\bibnamefont {Du{\v{s}}ek}}, \bibinfo {author} {\bibfnamefont {N.}~\bibnamefont {L{\"u}tkenhaus}},\ and\ \bibinfo {author} {\bibfnamefont {M.}~\bibnamefont {Peev}},\ }\bibfield  {title} {\bibinfo {title} {The security of practical quantum key distribution},\ }\href@noop {} {\bibfield  {journal} {\bibinfo  {journal} {Reviews of Modern Physics}\ }\textbf {\bibinfo {volume} {81}},\ \bibinfo {pages} {1301} (\bibinfo {year} {2009})}\BibitemShut {NoStop}%
\bibitem [{\citenamefont {Lo}\ \emph {et~al.}(2014)\citenamefont {Lo}, \citenamefont {Curty},\ and\ \citenamefont {Tamaki}}]{lo2014secure}%
  \BibitemOpen
  \bibfield  {author} {\bibinfo {author} {\bibfnamefont {H.-K.}\ \bibnamefont {Lo}}, \bibinfo {author} {\bibfnamefont {M.}~\bibnamefont {Curty}},\ and\ \bibinfo {author} {\bibfnamefont {K.}~\bibnamefont {Tamaki}},\ }\bibfield  {title} {\bibinfo {title} {Secure quantum key distribution},\ }\href@noop {} {\bibfield  {journal} {\bibinfo  {journal} {Nature Photonics}\ }\textbf {\bibinfo {volume} {8}},\ \bibinfo {pages} {595} (\bibinfo {year} {2014})}\BibitemShut {NoStop}%
\bibitem [{\citenamefont {Jain}\ \emph {et~al.}(2022)\citenamefont {Jain}, \citenamefont {Chin}, \citenamefont {Mani}, \citenamefont {Lupo}, \citenamefont {Nikolic}, \citenamefont {Kordts}, \citenamefont {Pirandola}, \citenamefont {Pedersen}, \citenamefont {Kolb}, \citenamefont {{\"O}mer} \emph {et~al.}}]{jain2022practical}%
  \BibitemOpen
  \bibfield  {author} {\bibinfo {author} {\bibfnamefont {N.}~\bibnamefont {Jain}}, \bibinfo {author} {\bibfnamefont {H.-M.}\ \bibnamefont {Chin}}, \bibinfo {author} {\bibfnamefont {H.}~\bibnamefont {Mani}}, \bibinfo {author} {\bibfnamefont {C.}~\bibnamefont {Lupo}}, \bibinfo {author} {\bibfnamefont {D.~S.}\ \bibnamefont {Nikolic}}, \bibinfo {author} {\bibfnamefont {A.}~\bibnamefont {Kordts}}, \bibinfo {author} {\bibfnamefont {S.}~\bibnamefont {Pirandola}}, \bibinfo {author} {\bibfnamefont {T.~B.}\ \bibnamefont {Pedersen}}, \bibinfo {author} {\bibfnamefont {M.}~\bibnamefont {Kolb}}, \bibinfo {author} {\bibfnamefont {B.}~\bibnamefont {{\"O}mer}}, \emph {et~al.},\ }\bibfield  {title} {\bibinfo {title} {Practical continuous-variable quantum key distribution with composable security},\ }\href@noop {} {\bibfield  {journal} {\bibinfo  {journal} {Nature communications}\ }\textbf {\bibinfo {volume} {13}},\ \bibinfo {pages} {4740} (\bibinfo {year} {2022})}\BibitemShut {NoStop}%
\bibitem [{\citenamefont {Zurek}(2003)}]{zurek2003decoherence}%
  \BibitemOpen
  \bibfield  {author} {\bibinfo {author} {\bibfnamefont {W.~H.}\ \bibnamefont {Zurek}},\ }\bibfield  {title} {\bibinfo {title} {Decoherence, einselection, and the quantum origins of the classical},\ }\href@noop {} {\bibfield  {journal} {\bibinfo  {journal} {Reviews of modern physics}\ }\textbf {\bibinfo {volume} {75}},\ \bibinfo {pages} {715} (\bibinfo {year} {2003})}\BibitemShut {NoStop}%
\bibitem [{\citenamefont {Henderson}\ and\ \citenamefont {Vedral}(2001)}]{henderson2001classical}%
  \BibitemOpen
  \bibfield  {author} {\bibinfo {author} {\bibfnamefont {L.}~\bibnamefont {Henderson}}\ and\ \bibinfo {author} {\bibfnamefont {V.}~\bibnamefont {Vedral}},\ }\bibfield  {title} {\bibinfo {title} {Classical, quantum and total correlations},\ }\href@noop {} {\bibfield  {journal} {\bibinfo  {journal} {Journal of physics A: mathematical and general}\ }\textbf {\bibinfo {volume} {34}},\ \bibinfo {pages} {6899} (\bibinfo {year} {2001})}\BibitemShut {NoStop}%
\bibitem [{\citenamefont {Portmann}\ and\ \citenamefont {Renner}(2022)}]{portmann2022security}%
  \BibitemOpen
  \bibfield  {author} {\bibinfo {author} {\bibfnamefont {C.}~\bibnamefont {Portmann}}\ and\ \bibinfo {author} {\bibfnamefont {R.}~\bibnamefont {Renner}},\ }\bibfield  {title} {\bibinfo {title} {Security in quantum cryptography},\ }\href@noop {} {\bibfield  {journal} {\bibinfo  {journal} {Reviews of Modern Physics}\ }\textbf {\bibinfo {volume} {94}},\ \bibinfo {pages} {025008} (\bibinfo {year} {2022})}\BibitemShut {NoStop}%
\bibitem [{\citenamefont {Zhang}\ \emph {et~al.}(2018)\citenamefont {Zhang}, \citenamefont {Xu}, \citenamefont {Chen}, \citenamefont {Peng},\ and\ \citenamefont {Pan}}]{zhang2018large}%
  \BibitemOpen
  \bibfield  {author} {\bibinfo {author} {\bibfnamefont {Q.}~\bibnamefont {Zhang}}, \bibinfo {author} {\bibfnamefont {F.}~\bibnamefont {Xu}}, \bibinfo {author} {\bibfnamefont {Y.-A.}\ \bibnamefont {Chen}}, \bibinfo {author} {\bibfnamefont {C.-Z.}\ \bibnamefont {Peng}},\ and\ \bibinfo {author} {\bibfnamefont {J.-W.}\ \bibnamefont {Pan}},\ }\bibfield  {title} {\bibinfo {title} {Large scale quantum key distribution: challenges and solutions},\ }\href@noop {} {\bibfield  {journal} {\bibinfo  {journal} {Optics express}\ }\textbf {\bibinfo {volume} {26}},\ \bibinfo {pages} {24260} (\bibinfo {year} {2018})}\BibitemShut {NoStop}%
\bibitem [{\citenamefont {Pirandola}\ \emph {et~al.}(2020)\citenamefont {Pirandola}, \citenamefont {Andersen}, \citenamefont {Banchi}, \citenamefont {Berta}, \citenamefont {Bunandar}, \citenamefont {Colbeck}, \citenamefont {Englund}, \citenamefont {Gehring}, \citenamefont {Lupo}, \citenamefont {Ottaviani} \emph {et~al.}}]{pirandola2020advances}%
  \BibitemOpen
  \bibfield  {author} {\bibinfo {author} {\bibfnamefont {S.}~\bibnamefont {Pirandola}}, \bibinfo {author} {\bibfnamefont {U.~L.}\ \bibnamefont {Andersen}}, \bibinfo {author} {\bibfnamefont {L.}~\bibnamefont {Banchi}}, \bibinfo {author} {\bibfnamefont {M.}~\bibnamefont {Berta}}, \bibinfo {author} {\bibfnamefont {D.}~\bibnamefont {Bunandar}}, \bibinfo {author} {\bibfnamefont {R.}~\bibnamefont {Colbeck}}, \bibinfo {author} {\bibfnamefont {D.}~\bibnamefont {Englund}}, \bibinfo {author} {\bibfnamefont {T.}~\bibnamefont {Gehring}}, \bibinfo {author} {\bibfnamefont {C.}~\bibnamefont {Lupo}}, \bibinfo {author} {\bibfnamefont {C.}~\bibnamefont {Ottaviani}}, \emph {et~al.},\ }\bibfield  {title} {\bibinfo {title} {Advances in quantum cryptography},\ }\href@noop {} {\bibfield  {journal} {\bibinfo  {journal} {Advances in Optics and Photonics}\ }\textbf {\bibinfo {volume} {12}},\ \bibinfo {pages} {1012} (\bibinfo {year} {2020})}\BibitemShut {NoStop}%
\bibitem [{\citenamefont {Sayat}\ \emph {et~al.}(2024)\citenamefont {Sayat}, \citenamefont {Shajilal}, \citenamefont {Kish}, \citenamefont {Assad}, \citenamefont {Symul}, \citenamefont {Lam}, \citenamefont {Rattenbury},\ and\ \citenamefont {Cater}}]{sayat2024satellite}%
  \BibitemOpen
  \bibfield  {author} {\bibinfo {author} {\bibfnamefont {M.~T.}\ \bibnamefont {Sayat}}, \bibinfo {author} {\bibfnamefont {B.}~\bibnamefont {Shajilal}}, \bibinfo {author} {\bibfnamefont {S.~P.}\ \bibnamefont {Kish}}, \bibinfo {author} {\bibfnamefont {S.~M.}\ \bibnamefont {Assad}}, \bibinfo {author} {\bibfnamefont {T.}~\bibnamefont {Symul}}, \bibinfo {author} {\bibfnamefont {P.~K.}\ \bibnamefont {Lam}}, \bibinfo {author} {\bibfnamefont {N.~J.}\ \bibnamefont {Rattenbury}},\ and\ \bibinfo {author} {\bibfnamefont {J.~E.}\ \bibnamefont {Cater}},\ }\bibfield  {title} {\bibinfo {title} {Satellite-to-ground continuous variable quantum key distribution: The gaussian and discrete modulated protocols in low earth orbit},\ }\href@noop {} {\bibfield  {journal} {\bibinfo  {journal} {IEEE Transactions on Communications}\ }\textbf {\bibinfo {volume} {72}},\ \bibinfo {pages} {3244} (\bibinfo {year} {2024})}\BibitemShut {NoStop}%
\bibitem [{\citenamefont {Usenko}\ \emph {et~al.}(2026)\citenamefont {Usenko}, \citenamefont {Ac{\'\i}n}, \citenamefont {All{\'e}aume}, \citenamefont {Andersen}, \citenamefont {Diamanti}, \citenamefont {Gehring}, \citenamefont {Hajomer}, \citenamefont {Kanitschar}, \citenamefont {Pacher}, \citenamefont {Pirandola} \emph {et~al.}}]{usenko2026continuous}%
  \BibitemOpen
  \bibfield  {author} {\bibinfo {author} {\bibfnamefont {V.~C.}\ \bibnamefont {Usenko}}, \bibinfo {author} {\bibfnamefont {A.}~\bibnamefont {Ac{\'\i}n}}, \bibinfo {author} {\bibfnamefont {R.}~\bibnamefont {All{\'e}aume}}, \bibinfo {author} {\bibfnamefont {U.~L.}\ \bibnamefont {Andersen}}, \bibinfo {author} {\bibfnamefont {E.}~\bibnamefont {Diamanti}}, \bibinfo {author} {\bibfnamefont {T.}~\bibnamefont {Gehring}}, \bibinfo {author} {\bibfnamefont {A.~A.}\ \bibnamefont {Hajomer}}, \bibinfo {author} {\bibfnamefont {F.}~\bibnamefont {Kanitschar}}, \bibinfo {author} {\bibfnamefont {C.}~\bibnamefont {Pacher}}, \bibinfo {author} {\bibfnamefont {S.}~\bibnamefont {Pirandola}}, \emph {et~al.},\ }\bibfield  {title} {\bibinfo {title} {Continuous-variable quantum communication},\ }\href@noop {} {\bibfield  {journal} {\bibinfo  {journal} {Reviews of Modern Physics}\ }\textbf {\bibinfo {volume} {98}},\ \bibinfo {pages} {015003} (\bibinfo {year} {2026})}\BibitemShut {NoStop}%
\bibitem [{\citenamefont {Ra}\ \emph {et~al.}(2020)\citenamefont {Ra}, \citenamefont {Dufour}, \citenamefont {Walschaers}, \citenamefont {Jacquard}, \citenamefont {Michel}, \citenamefont {Fabre},\ and\ \citenamefont {Treps}}]{ra2020non}%
  \BibitemOpen
  \bibfield  {author} {\bibinfo {author} {\bibfnamefont {Y.-S.}\ \bibnamefont {Ra}}, \bibinfo {author} {\bibfnamefont {A.}~\bibnamefont {Dufour}}, \bibinfo {author} {\bibfnamefont {M.}~\bibnamefont {Walschaers}}, \bibinfo {author} {\bibfnamefont {C.}~\bibnamefont {Jacquard}}, \bibinfo {author} {\bibfnamefont {T.}~\bibnamefont {Michel}}, \bibinfo {author} {\bibfnamefont {C.}~\bibnamefont {Fabre}},\ and\ \bibinfo {author} {\bibfnamefont {N.}~\bibnamefont {Treps}},\ }\bibfield  {title} {\bibinfo {title} {Non-gaussian quantum states of a multimode light field},\ }\href@noop {} {\bibfield  {journal} {\bibinfo  {journal} {Nature Physics}\ }\textbf {\bibinfo {volume} {16}},\ \bibinfo {pages} {144} (\bibinfo {year} {2020})}\BibitemShut {NoStop}%
\bibitem [{\citenamefont {Lvovsky}\ \emph {et~al.}(2020)\citenamefont {Lvovsky}, \citenamefont {Grangier}, \citenamefont {Ourjoumtsev}, \citenamefont {Parigi}, \citenamefont {Sasaki},\ and\ \citenamefont {Tualle-Brouri}}]{lvovsky2020production}%
  \BibitemOpen
  \bibfield  {author} {\bibinfo {author} {\bibfnamefont {A.}~\bibnamefont {Lvovsky}}, \bibinfo {author} {\bibfnamefont {P.}~\bibnamefont {Grangier}}, \bibinfo {author} {\bibfnamefont {A.}~\bibnamefont {Ourjoumtsev}}, \bibinfo {author} {\bibfnamefont {V.}~\bibnamefont {Parigi}}, \bibinfo {author} {\bibfnamefont {M.}~\bibnamefont {Sasaki}},\ and\ \bibinfo {author} {\bibfnamefont {R.}~\bibnamefont {Tualle-Brouri}},\ }\bibfield  {title} {\bibinfo {title} {Production and applications of non-gaussian quantum states of light},\ }\href@noop {} {\bibfield  {journal} {\bibinfo  {journal} {arXiv preprint arXiv:2006.16985}\ } (\bibinfo {year} {2020})}\BibitemShut {NoStop}%
\bibitem [{\citenamefont {Jeng}\ \emph {et~al.}(2025)\citenamefont {Jeng}, \citenamefont {Lam},\ and\ \citenamefont {Assad}}]{jeng2025entanglement}%
  \BibitemOpen
  \bibfield  {author} {\bibinfo {author} {\bibfnamefont {H.}~\bibnamefont {Jeng}}, \bibinfo {author} {\bibfnamefont {P.~K.}\ \bibnamefont {Lam}},\ and\ \bibinfo {author} {\bibfnamefont {S.~M.}\ \bibnamefont {Assad}},\ }\bibfield  {title} {\bibinfo {title} {Entanglement-based quantum key distribution with non-gaussian continuous variables},\ }\href@noop {} {\bibfield  {journal} {\bibinfo  {journal} {Proceedings of the Royal Society A}\ }\textbf {\bibinfo {volume} {481}},\ \bibinfo {pages} {20240970} (\bibinfo {year} {2025})}\BibitemShut {NoStop}%
\bibitem [{\citenamefont {Straka}\ \emph {et~al.}(2018)\citenamefont {Straka}, \citenamefont {Lachman}, \citenamefont {Hlou{\v{s}}ek}, \citenamefont {Mikov{\'a}}, \citenamefont {Mi{\v{c}}uda}, \citenamefont {Je{\v{z}}ek},\ and\ \citenamefont {Filip}}]{straka2018quantum}%
  \BibitemOpen
  \bibfield  {author} {\bibinfo {author} {\bibfnamefont {I.}~\bibnamefont {Straka}}, \bibinfo {author} {\bibfnamefont {L.}~\bibnamefont {Lachman}}, \bibinfo {author} {\bibfnamefont {J.}~\bibnamefont {Hlou{\v{s}}ek}}, \bibinfo {author} {\bibfnamefont {M.}~\bibnamefont {Mikov{\'a}}}, \bibinfo {author} {\bibfnamefont {M.}~\bibnamefont {Mi{\v{c}}uda}}, \bibinfo {author} {\bibfnamefont {M.}~\bibnamefont {Je{\v{z}}ek}},\ and\ \bibinfo {author} {\bibfnamefont {R.}~\bibnamefont {Filip}},\ }\bibfield  {title} {\bibinfo {title} {Quantum non-gaussian multiphoton light},\ }\href@noop {} {\bibfield  {journal} {\bibinfo  {journal} {npj Quantum Information}\ }\textbf {\bibinfo {volume} {4}},\ \bibinfo {pages} {4} (\bibinfo {year} {2018})}\BibitemShut {NoStop}%
\bibitem [{\citenamefont {Nagaraja~Rao}\ \emph {et~al.}(2026)\citenamefont {Nagaraja~Rao}, \citenamefont {Murphy}, \citenamefont {Ahwal}, \citenamefont {Medlock}, \citenamefont {Spiller},\ and\ \citenamefont {Kumar}}]{nagaraja2026quantum}%
  \BibitemOpen
  \bibfield  {author} {\bibinfo {author} {\bibfnamefont {V.}~\bibnamefont {Nagaraja~Rao}}, \bibinfo {author} {\bibfnamefont {K.}~\bibnamefont {Murphy}}, \bibinfo {author} {\bibfnamefont {F.}~\bibnamefont {Ahwal}}, \bibinfo {author} {\bibfnamefont {E.~T.~H.}\ \bibnamefont {Medlock}}, \bibinfo {author} {\bibfnamefont {T.~P.}\ \bibnamefont {Spiller}},\ and\ \bibinfo {author} {\bibfnamefont {R.}~\bibnamefont {Kumar}},\ }\bibfield  {title} {\bibinfo {title} {Quantum random number generation from the continuous variable payload for the spoqc mission},\ }\href@noop {} {\bibfield  {journal} {\bibinfo  {journal} {Quantum Science and Technology}\ } (\bibinfo {year} {2026})}\BibitemShut {NoStop}%
\bibitem [{\citenamefont {Vokes}\ \emph {et~al.}(2026)\citenamefont {Vokes}, \citenamefont {Rao}, \citenamefont {Spencer},\ and\ \citenamefont {Kumar}}]{vokes2026energy}%
  \BibitemOpen
  \bibfield  {author} {\bibinfo {author} {\bibfnamefont {E.}~\bibnamefont {Vokes}}, \bibinfo {author} {\bibfnamefont {V.~N.}\ \bibnamefont {Rao}}, \bibinfo {author} {\bibfnamefont {E.}~\bibnamefont {Spencer}},\ and\ \bibinfo {author} {\bibfnamefont {R.}~\bibnamefont {Kumar}},\ }\bibfield  {title} {\bibinfo {title} {Energy efficient optical tracking for space quantum communication},\ }\href@noop {} {\bibfield  {journal} {\bibinfo  {journal} {arXiv preprint arXiv:2602.22160}\ } (\bibinfo {year} {2026})}\BibitemShut {NoStop}%
\bibitem [{\citenamefont {Rao}\ \emph {et~al.}(2025)\citenamefont {Rao}, \citenamefont {Medlock}, \citenamefont {Spiller},\ and\ \citenamefont {Kumar}}]{rao2025continuous}%
  \BibitemOpen
  \bibfield  {author} {\bibinfo {author} {\bibfnamefont {V.~N.}\ \bibnamefont {Rao}}, \bibinfo {author} {\bibfnamefont {E.~T.~H.}\ \bibnamefont {Medlock}}, \bibinfo {author} {\bibfnamefont {T.}~\bibnamefont {Spiller}},\ and\ \bibinfo {author} {\bibfnamefont {R.}~\bibnamefont {Kumar}},\ }\bibfield  {title} {\bibinfo {title} {Continuous-variable quantum key distribution with a single quadrature measurement at an arbitrary reference frame},\ }\href@noop {} {\bibfield  {journal} {\bibinfo  {journal} {Physical Review A}\ }\textbf {\bibinfo {volume} {112}},\ \bibinfo {pages} {042406} (\bibinfo {year} {2025})}\BibitemShut {NoStop}%
\end{thebibliography}%

\appendix

\section{Preparation of GMCS}

A coherent state is defined by
\begin{equation}
\ket{\alpha} = \hat{D}(\alpha)\ket{0},
\end{equation}
where $\hat{D}(\alpha) = \exp(\alpha \hat{a}^\dagger - \alpha^* \hat{a})$ is the displacement operator, $\hat{a}$ and $\hat{a}^\dagger$ are the annihilation and creation operators, and $\ket{0}$ denotes the vacuum state. The complex amplitude $\alpha$ is related to the quadratures $(x_A, p_A)$ by
\begin{equation}
\alpha = \frac{1}{\sqrt{2}}(x_A + i p_A).
\end{equation}

CVP generates two independent real valued random variables $x_A$ and $p_A$, drawn from a Gaussian distribution as
\begin{equation}
x_A, p_A \sim \mathcal{N}(0, V_A),
\end{equation}
where $V_A$ is the modulation variance (in SNU). The prepared coherent state is thus $\ket{\alpha(x_A, p_A)}$.

The electrical signals applied to the modulators are generated from digital-to-analog converters (DACs), which output voltages proportional to the sampled Gaussian variables $x_A$ and $p_A$. These voltages induce proportional changes in the optical field quadratures, thereby implementing the displacement operator $\hat{D}(\alpha)$.

The resulting optical field can be expressed in terms of quadrature operators as
\begin{equation}
\hat{x} = x_A + \hat{x}_0, \quad
\hat{p} = p_A + \hat{p}_0,
\end{equation}
where $\hat{x}_0$ and $\hat{p}_0$ represent the vacuum fluctuations with variance normalized to one shot noise unit.

Over many pulses, CVP prepares an ensemble of coherent states with a Gaussian distribution in phase space. The corresponding density operator is given by
\begin{widetext}
\begin{equation}
\rho = \int \frac{1}{2\pi V_A} 
\exp\left(-\frac{x_A^2 + p_A^2}{2V_A}\right)
\ket{\alpha(x_A,p_A)}\bra{\alpha(x_A,p_A)} \, dx_A \, dp_A.
\end{equation}
\end{widetext}

This ensemble forms a Gaussian modulated distribution in phase space, which is optimal for the CV-QKD protocol under collective Gaussian attacks. The output of this is essentially plotted in Fig. \ref{fig:Gaussian_state}, with the discretization of the same given in Fig. \ref{fig:GM_pattern}.

\section{CV-QRNG}

A CV-QRNG based on vacuum noise detection employs a shot noise limited homodyne detector, where the measured quadrature contains contributions from both quantum and classical noise sources. However, this decomposition has inherent limitations in accurately quantifying the true quantum randomness. While the entropy contributions from classical noise sources can be reasonably estimated, the quantum component cannot be directly isolated with complete certainty. In a conservative security framework, it is therefore assumed that an eavesdropper (Eve) has full knowledge of the classical noise contributions and may also possess partial information about the quantum component of the entropy. The details of the experimental methodology are as follows. To extract potential randomness from vacuum states, datasets are acquired under three different configurations:
\begin{enumerate}
\item The optical power of the local oscillator (LO) is varied to estimate $H(X)_{c(1)}$, corresponding to noise arising from imperfect homodyne detection.
\item The LO is switched off while all electrical components remain active to estimate $H(X)_{c(2)}$, representing the electronic noise contribution.
\item Both the LO and electrical components are active to estimate $H(X)_{\text{total}}$, corresponding to the total measured noise, including shot noise.
\end{enumerate}

Among the various sources contributing to classical entropy, only $H(X)_{c(2)}$ is observed to exhibit dynamic behavior, while the remaining noise sources are assumed to remain constant for a fixed temperature and operating condition. This assumption is justified by the negligible variation in parameters such as beam-splitter reflectivity and detector inefficiencies (Fig.~\ref{fig:design}). However, environmental factors, particularly temperature and pressure, may influence the performance of these components. Therefore, the QRNG is operated only within a specified range of temperature and pressure over which the system has been experimentally characterized. A sample of the raw data of the shot noise and electronic noise for QRNG are given in Fig. \ref{fig:Raw_data}.

Assuming that the classical noise components are independent, the Shannon entropy associated with the quantum contribution, $H(X)_q$, can be expressed as
\begin{equation}
H(X)_q = H(X)_{\text{total}} - H(X)_{c(i)},
\label{eq:entropy}
\end{equation}
where the Shannon entropy is defined as $H(X) = -\sum_i p_i \log_2 p_i$, with $p_i$ denoting the probability distribution of the variable $X$, and $c(i)$ representing the different classical noise contributions.

\begin{figure}[h]
\centering
\includegraphics[width=0.95\linewidth]{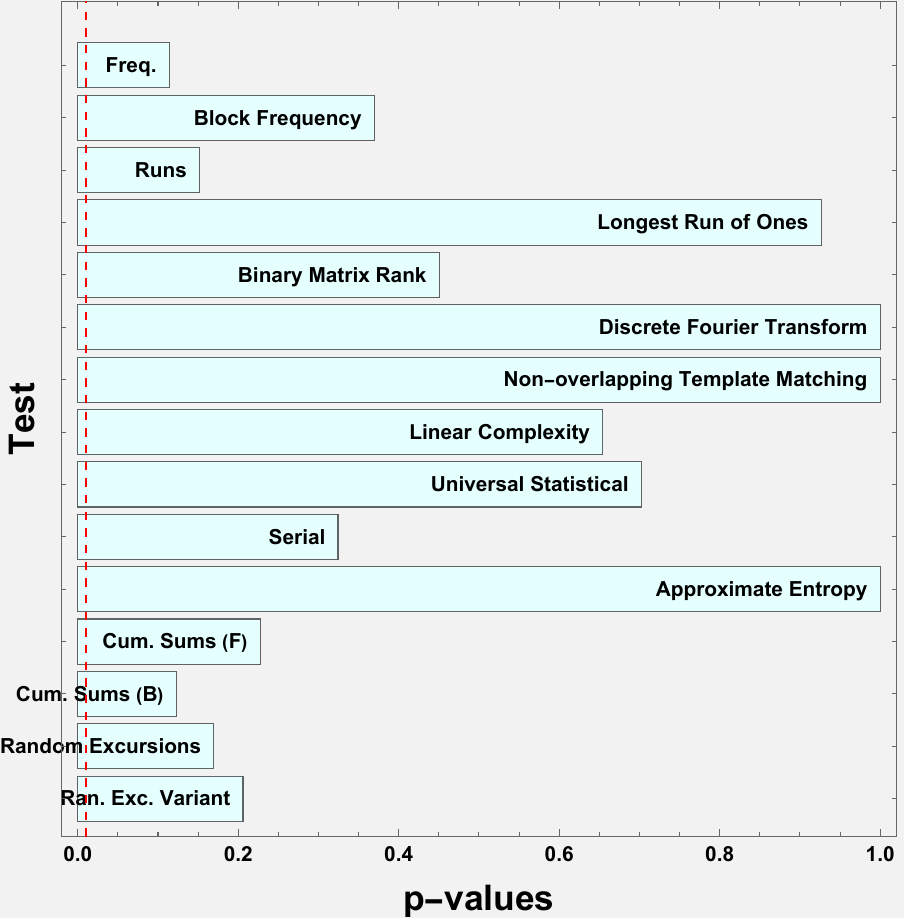}
\caption{The output of the NIST test suite from CV-QRNG. The minimum p-value for the tests is given in red dashed line.}
\label{fig:qrng}
\end{figure}

\end{document}